%% file: main.tex
\documentclass[10pt,a4paper,twocolumn]{article}
\usepackage{f1000_styles}
\usepackage[round]{natbib}
\let\cite\citep

\usepackage[english]{babel}
\numberwithin{equation}{section}

\usepackage{amsmath, amssymb}
\usepackage{bbm}
\usepackage{graphicx}
\usepackage[colorlinks=true, allcolors=blue]{hyperref}
\usepackage{dsfont}
\usepackage{csquotes}
\usepackage{booktabs, colortbl}
\usepackage{wrapfig}
\usepackage{pgfplotstable}
\usepackage{makecell}
\pgfplotsset{compat=1.14}
\pgfplotstableset{fixed, fixed zerofill, precision = 3}
\usepackage{listings}
\usepackage{color}

\definecolor{dkgreen}{rgb}{0,0.6,0}
\definecolor{gray}{rgb}{0.5,0.5,0.5}
\definecolor{mauve}{rgb}{0.58,0,0.82}

\newcommand{\cA}{\mathcal A}
\newcommand{\cE}{\mathcal E}
\newcommand{\cF}{\mathcal F}
\newcommand{\cR}{\mathcal R}
\newcommand{\lingform}[1]{{\it #1}}
\DeclareMathOperator{\ExpNull}{\mathbf{E}_0}

\DeclareMathOperator{\Exp}{\mathbf{E}}
\DeclareMathOperator{\Var}{\mathbf{Var}}
\DeclareMathOperator{\ProbNull}{\mathbf{P}_0}
\DeclareMathOperator{\ProbAlt}{\mathbf{P}_1}
\DeclareMathOperator{\Prob}{\mathbf{P}}
\DeclareMathOperator{\LR}{\mathbf{LR}\scalebox{0.6}[0.6]{\textsc{10}}}

\newcommand{\minus}{\scalebox{0.75}[1.0]{$-$}}
\providecommand{\abs}[1]{\lvert#1\rvert}

\begin{document}

\title{Accumulation Bias in meta-analysis: \\ the need to consider \lingform{time} in error control}
\author[1]{Judith ter Schure}
\author[2]{Peter Gr\"unwald}
\affil[1]{CWI, Science Park 123, 1098 XG Amsterdam, The Netherlands, schure@cwi.nl (corresponding author)}
\affil[2]{CWI, Science Park 123, 1098 XG Amsterdam, The Netherlands, pdg@cwi.nl}

\maketitle
\thispagestyle{fancy}

\begin{abstract}
Studies accumulate over time and meta-analyses are mainly retrospective. These two characteristics introduce dependencies between the \lingform{analysis time}, at which a series of studies is up for meta-analysis, and results within the series. Dependencies introduce bias --- \lingform{Accumulation Bias} --- and invalidate the sampling distribution assumed for p-value tests, thus inflating type-I errors. But dependencies are also inevitable, since for science to accumulate efficiently, new research needs to be informed by past results. Here, we investigate various ways in which \lingform{time} influences error control in meta-analysis testing. We introduce an \lingform{Accumulation Bias Framework} that allows us to model a wide variety of practically occurring dependencies, including study series accumulation, meta-analysis timing, and approaches to multiple testing in living systematic reviews. The strength of this framework is that it shows how all dependencies affect p-value-based tests in a similar manner. This leads to two main conclusions. First, Accumulation Bias is inevitable, and even if it can be approximated and accounted for, no valid p-value tests can be constructed. Second, tests based on likelihood ratios withstand Accumulation Bias: they provide bounds on error probabilities that remain valid despite the bias. We leave the reader with a choice between two proposals to consider \lingform{time} in error control: either treat individual (primary) studies and meta-analyses as two separate worlds --- each with their own timing --- or integrate individual studies in the meta-analysis world. Taking up likelihood ratios in either approach allows for valid tests that relate well to the accumulating nature of scientific knowledge. Likelihood ratios can be interpreted as betting profits, earned in previous studies and invested in new ones, while the meta-analyst is allowed to cash out at any time and advise against future studies.
\end{abstract}

\section*{Keywords}
meta-analysis, accumulation bias, sequential, cumulative, living systematic review, likelihood ratio, research waste, evidence-based research

\clearpage

\section{Introduction}
\begin{quote}
Meta-analysis refers to the statistical synthesis of results from a series of studies. [...] the synthesis will be meaningful only if the studies have been collected systematically. [...] The formulas used in meta-analysis are extensions of formulas used in primary studies, and are used to address similar kinds of questions to those addressed in primary studies.
--Borenstein, Hedges, Higgins \& Rothstein (\citeyear[pp. xxi-xxiii]{borenstein2009introduction})
\end{quote}
\begin{quote}
To consult the statistician after an experiment is finished is often merely to ask him to conduct a post mortem examination. He can perhaps say what the experiment died of.
--Fisher (\citeyear[p. 18]{fisher1938presidential})
\end{quote}
These two quotes conflict. Most meta-analyses are retrospective and consider the number of studies available --- after the literature has been searched systematically --- as a given for the statistical analysis. P-value based statistical tests, however, are intended to be prospective and require the sample size --- or the stopping rule that produces the sample --- to be set specifically for the planned statistical analysis. The second quote, by the p-value's popularizer Ronald Fisher, is about primary studies. But this prospective rationale influences meta-analysis as well because it also involves the size of the study series: p-value tests assume that the number of studies --- so the timing of the meta-analysis --- is predetermined or at least unrelated to the study results. So by using p-value methods, conventional meta-analysis implicitly assumes that promising initial results are just as likely to develop into (large) series of studies as their disappointing counterparts. Conclusive studies should just as likely trigger meta-analyses as inconclusive ones. And so the use of p-value tests suggests that results of earlier studies should be unknown when planning new studies as well as when planning meta-analyses. Such assumptions are unrealistic and actively argued against by the \lingform{Evidence-Based Research Network} \citep{lund2016towards} part of the movement to reduce research waste \citep{chalmers2009avoidable, chalmers2014increase}. But ignoring these assumptions invalidates conventional p-value tests and inflates type-I errors.

P-values are based on tail areas of a test statistic's sampling distribution under the null hypothesis, and thus require this distribution to be fully specified. In this paper we show that the standard normal \(Z\)-distribution generally assumed (e.g. \citet{borenstein2009introduction}) is not an appropriate sampling distribution. Moreover, we believe that no sampling distribution can be specified that fully represents the variety of processes in accumulating scientific knowledge and all decision made along the way. We need a more flexible approach to testing that controls errors regardless of the process that spurs the meta-analysis.

When dependencies arise between study series size or meta-analysis timing and results within the series, bias is introduced in the estimates. This bias is inherent to accumulating data, which is why we gave it the name \lingform{Accumulation Bias}. Various forms of Accumulation Bias have been characterized before, in very general terms as \enquote{bias introduced by the order in which studies are conducted} \citep[p. 197]{whitehead2002meta} and more specifically, such as bias caused by the dependence of follow-up studies on previous studies' significance and the dependence of meta-analysis timing on previous study results \citep{ellis2009temporal}. Also, more elaborate relations were studied between the existence of follow-up studies, study design and meta-analysis estimates \citep{kulinskaya2016sequential}. Yet no approach to confront these biases has been proposed.

In this paper we define \lingform{Accumulation Bias} to encompass processes that not only affect parameter estimates but also the shape of the sampling distribution, which is why only approximation and correction for bias does not achieve valid p-value tests. We illustrate this by an example in Section \ref{sec:GoldRush}, right after we give a general introduction to Accumulation Bias in Section \ref{sec:AccumBias} with its relation to publication bias (Section \ref{sec:PubBias}) and an informal characterization of the direction of the bias (Section \ref{sec:Direction}). By presenting its diversity, we argue throughout the paper that any efficient scientific process will introduce some form of Accumulation Bias and that the exact process can never be fully known. We collect the various forms of Accumulation Bias into one framework (Section \ref{sec:Framework}) and show that all are related to the \lingform{time} aspect in meta-analysis. The framework incorporates dependencies mentioned by \citet{whitehead2002meta}, \citet{ellis2009temporal} and \citet{kulinskaya2016sequential} as well the effect of multiple testing over time in living systematic reviews \cite{simmonds2017living}. We conclude that some version of these biases will also be introduced by \lingform{Evidence-Based Research}.

Our framework specifies \lingform{analysis time probabilities} --- with behavior familiar from survival analysis --- and distinguishes two approaches to error control: conditional on time (Section \ref{sec:Cond}) and surviving over time (Section \ref{sec:Surv}). We show that general meta-analyses take the former approach, while existing methods for living systematic reviews take the latter. However, neither of the two is able to analyze study series affected by partially unknown processes of Accumulation Bias (Section \ref{sec:Unknown}). After an intermezzo on evidence that indeed such processes are already at play in Section \ref{sec:Evid}, we introduce a general form of a test statistic that is able to withstand any Accumulation Bias process: the likelihood ratio. We specify bounds on error probabilities that are valid despite the existing bias, for error control conditional on time (Section \ref{sec:LRcondition}) as well as surviving over time (Section \ref{sec:LRsurviving}). The reader is left to choose between the two; the consequences of either preference are specified in Section \ref{sec:Choice}. We try to give intuition on why both are still possible in their respective sections \ref{sec:LRcondition} and \ref{sec:LRsurviving}, but also give some extra intuition on the magic of likelihood ratios in Section \ref{sec:Betting}: Likelihood ratios have an interpretation as betting profit that can be reinvested in future studies. At the same time, the meta-analyst is allowed to cash out at any time and advise against future studies. Hence, the likelihood ratio relates the statistics of Accumulation Bias to the accumulating nature of scientific knowledge, which is critical in reducing research waste.

\section{Accumulation Bias} \label{sec:AccumBias}

Any meta-analyst carries out a meta-analysis under the assumption that synthesizing previous studies will add to what is already known from existing studies. So meta-analyses are mainly performed on series of studies of meaningful size. What is considered meaningful varies considerably: 16 and 15 studies per meta-analysis were reported to be the median numbers in \lingform{Medline} meta-analyses from 2004 and 2014 \citep{moher2007epidemiology, page2016epidemiology}, while 3 studies per meta-analysis were reported in \lingform{Cochrane} meta-analyses from 2008 (\lingform{Cochrane Database of Systematic Reviews} \cite{davey2011characteristics}). Since meta-analyses are performed on research hypotheses that have spurred a certain study series size, they always report estimates that are conditioned on the availability of such a series. The crucial point is that not all pilot studies or small study series will reach a meaningful size, and that doing so might depend on results in the series. Apart from the dependent size of the study series, the exact timing of a meta-analysis can also depend on the available results. The completion of a highly powered or otherwise conclusive study, for example, might be considered to finalize the series and trigger a meta-analysis. So meta-analysis also report estimates conditioned on the consideration that a systematic synthesis will be informative. Both dependencies --- series size and meta-analysis timing --- introduce bias: Accumulation Bias.

\subsection{Accumulation Bias vs. publication bias} \label{sec:PubBias}

Publication bias refers to the practice that studies with nonsignificant, or more general, unsatisfactory results have smaller probability to be published than studies with significant, satisfactory results. So unsatisfactory studies are performed, but do not reach the meta-analyst because they are stashed away in a file drawer \citep{rosenthal1979file}. Accumulation Bias, on the other hand, refers to some studies or meta-analyses not being performed at all, as a result of previous findings in a series of studies. In a file drawer-free world, Accumulation Bias would still exist. But Accumulation Bias is a manageable problem because it does not operate at the individual study level. Conditional on the fact that a second study is performed, the second study is an unbiased sample. Conditional on the fact that a third study is performed, for whatever reason, the third study is an unbiased sample. So bias is introduced at the level of the series, not at the study level. This is different for publication bias, where, conditional on being published, the studies available are not an unbiased sample. We exploit the difference in this paper by considering \lingform{time} in error control.

Of course, Accumulation Bias and publication bias are not alone in their effects on meta-analysis reporting. All sorts of \lingform{significance chasing biases} --- selective-outcome bias, selective analysis reporting bias and fabrication bias --- might be present in the study series up for meta-analysis, and can lead to \enquote{wrong and misleading answers} \citep[p. 169]{ioannidis2010meta}. But for a world in which these biases are overcome, we also need tests that reflect how scientific knowledge accumulates.

\subsection{Accumulation Bias' direction} \label{sec:Direction}

Accumulation Bias in estimates is mainly bias in the satisfactory direction, which means that the effect under study is overestimated. This is the case for bias caused by size of the studies series when (overly) optimistic initial estimates (either in individual studies or in intermediate meta-analyses) give rise to more studies, while disappointing results terminate a series of studies. This is also the case when the timing of the meta-analysis is based on an (overly) optimistic last study estimate or an (overly) optimistic meta-analysis synthesis is considered the final one. We focus on this satisfactory direction of Accumulation Bias and will only briefly discuss other possibilities in Section \ref{sec:Unknown} and \ref{sec:EmpEvid}. We introduce the wide variety of possible dependencies in an \lingform{Accumulation Bias Framework} in Section \ref{sec:Framework}, which has a generality that also includes Accumulation Bias without a clear direction. But we first present Accumulation Bias' effects on error control by an example.

\section{A \lingform{Gold Rush} example: new studies after finding significant results} \label{sec:GoldRush}
We study the effect of Accumulation Bias by a simple example. Its simplicity allows us to calculate the exact amount of bias in the test statistic and investigate the additional effect on the sampling distribution. The example given in this section is an extension of the toy example introduced by \citet{ellis2009temporal}. We denote this example by \lingform{Gold Rush} because it describes how new studies go looking for more results after finding initial statistical significance. In the current culture of scientific practice, statistical significance can be seen as the currency of scientific success. After all, significant results achieve the future possibility to pay off in publications, grants and tenure positions. When a gold rush for statistical significance presents itself in a series of studies, dependencies arise between the size of the series and the results within: Accumulation Bias. We specify this mechanism in detail in Section \ref{sec:GoldNewStudyProbs} and \ref{sec:GoldNewStudyProbsInd}, after we simplified our meta-analysis setting to common/fixed-effects meta-analysis in Section \ref{sec:GoldCommFix}. We present the resulting bias in the test estimates in Section \ref{sec:GoldEstimates} and its additional effects on the sampling distribution and testing in Section \ref{sec:GoldDistribution} and \ref{sec:GoldTesting}. In Section \ref{sec:GoldOccur} we conclude by pointing out the very mild condition needed for some form of \lingform{Gold Rush} Accumulation Bias to occur

\subsection{Common/fixed-effect meta-analysis} \label{sec:GoldCommFix}

This paper discusses meta-analysis in its simplest form, which is common-effect meta-analysis, also known as fixed-effect meta-analysis. This restriction does not mean that more complex forms of meta-analysis, such as random-effects meta-analysis and meta-regression, do not suffer from the problems mentioned in this paper. The reason for simplification is to reduce the complexity in quantifying the problem, part of showing that quantification is not enough. In a future paper we will study the effects of heterogeneity on testing in more detail. For an example of Accumulation Bias in random-effects estimates we refer to \citet{kulinskaya2016sequential}. 

Common-effect meta-analysis derives a combined \(Z\)-score from the summary statistics of the available studies. This combined \(Z\)-score is used as a test statistic in two-sided meta-analysis testing by comparing it to the tails of a standard normal distribution. This is equivalent to assessing whether its absolute value is more than \(z_{\frac{\alpha}{2}}\) standard deviations away from zero (larger than 1.960 for \(\alpha = 0.05\)). We simplify the setting by assuming studies with equal standard deviations to obtain an easy to handle expression for the combined \(Z\)-score of \(t\) available studies. We denote this meta-analysis \(Z\)-score by \(Z^{(t)}\) and derive it as the weighted average over the study \(Z\)-scores \(Z_1, \ldots, Z_t \), shown in its general form in Eq. \eqref{Zt:Unequaln} and in Eq. \eqref{Zt:Equaln} under the assumption of equal study sizes:

\begin{subequations}
\begin{align}
\label{Zt:Unequaln}
Z^{(t)} &= \frac{\sum_{i = 1}^{t} \sqrt{n_{i}}Z_{i}}{\sqrt{N^{(t)}}}
\quad \text{with} \quad N^{(t)} = \sum_{i = 1}^{t} n_{i} \\
\label{Zt:Equaln}
&= \frac{1}{\sqrt{t}} \sum_{i = 1}^t Z_i \quad (n_1 = n_2 = \dots = n_t = n).
\end{align}
\end{subequations}
See Appendix \ref{sec:AppCommFixMetaAnalysis} for a derivation from the mean difference notation in \citet{borenstein2009introduction}.

\subsection{\lingform{Gold Rush} new study probabilities} \label{sec:GoldNewStudyProbs}

In our \lingform{Gold Rush} example, we assume the following dependency within a series of studies: each study in a series has a larger probability to be replicated --- and therefore expanding the series of studies --- if the study shows a significant positive effect. So the existence of a new study is dependent on the significance and sign of the results of its predecessor.

\(T\) is the random variable that denotes the maximum size of a study series --- the time at which the search stops. We enumerate time by the order of appearance in a study series, with \(t = 1\) for the pilot study, \(t = 2\) for the second study (so now we have a two-study series) etc. So we use \(t\) to denote the number of studies available for meta-analysis at any time point: our notion of time is not related to actual dates at which studies are performed. The maximum time \(T\) is usually unknown since more studies might be performed in the future. \(T \geq 2\) means that the series has not halted after the first initial study, but that it is unknown how many replications will eventually be performed. In our extended \lingform{Gold Rush} example, we present the Accumulation Bias process by the probability that the maximum size is at least one study larger than the current size (\(T \geq t + 1\)), and do so using six parameters. We denote these parameters by the \lingform{new study probabilities}, since they indicate the probability that a follow-up study is performed when the result of the current study is available:

\begin{align}
\nonumber
\omega_{\textsc{s}}^{(1)} :=& \Prob\left[T \geq 2\,\middle|\, T \geq 1, Z_1 \geq z_{\frac{\alpha}{2}} \right] &= \quad &1\\
\nonumber
\omega_{\textsc{x}}^{(1)} :=& \Prob\left[T \geq 2\,\middle|\, T \geq 1, Z_1 \leq \minus z_{\frac{\alpha}{2}} \right] &= \quad &0\\
\nonumber
\omega_{\textsc{ns}}^{(1)} :=& \Prob\left[T \geq 2\,\middle|\, T \geq 1,\abs{Z_1} < z_{\frac{\alpha}{2}} \right] &= \quad &0.1,
\end{align}
\begin{align} \label{newStudyProbs}
\nonumber \\
\text{for all } t \geq 2: \\
\nonumber
\omega_{\textsc{s}}^{(t)} = \,\, \omega_{\textsc{s}} :=& \, \Prob\left[T \geq t + 1 \,\middle|\, T \geq t, Z_t \geq z_{\frac{\alpha}{2}} \right] &= \quad &1\\
\nonumber
\omega_{\textsc{x}}^{(t)} = \,\, \omega_{\textsc{x}} :=& \, \Prob\left[T \geq t + 1 \,\middle|\, T \geq t, Z_t \leq \minus z_{\frac{\alpha}{2}} \right] &= \quad &0\\
\nonumber
\omega_{\textsc{ns}}^{(t)} = \omega_{\textsc{ns}} :=& \, \Prob\left[T \geq t + 1 \,\middle|\, T \geq t, \abs{Z_t} < z_{\frac{\alpha}{2}} \right] &= \quad &0.02.
\end{align}
We distinguish between the influence of the first pilot study (\(\omega_{\textsc{s}}^{(1)}\), \(\omega_{\textsc{x}}^{(1)}\) and \(\omega_{\textsc{ns}}^{(1)}\)) and the others (\(\omega_{\textsc{s}}\), \(\omega_{\textsc{x}}\) and \(\omega_{\textsc{ns}}\)) since pilot studies are carried out with future studies in mind, and therefore replications have higher probability after the first than after other studies in the series, also in case the pilot study is not significant. We assume that no new study is performed when a significant negative result is obtained (\(\omega_{\textsc{x}}^{(1)} = \omega_{\textsc{x}} = 0\)) and new studies are always performed after positive significant findings, the satisfactory result (\(\omega_{\textsc{s}}^{(1)} = \omega_{\textsc{s}} = 1\)). Nonsignificant results have a small, but not negligible probability to spur new studies (\(\omega_{\textsc{ns}}^{(1)} = 0.1\), \(\omega_{\textsc{ns}} = 0.02\)).

\subsection{\lingform{Gold Rush} new study probabilities' independence from data-generating hypothesis} \label{sec:GoldNewStudyProbsInd}

In the following we use \(\ProbAlt\) to express probabilities under the alternative hypothesis and \(\ProbNull\) to express probabilities under the null hypothesis. Our new study probabilities in \eqref{newStudyProbs} were given without reference to any of these hypotheses, to make explicit that they depend solely on the data (or summary statistic \(Z_t\)) and not on the hypothesis that generated the data. So \(\Prob\) in these definitions can be read as \(\ProbAlt\) as well as \(\ProbNull\).

\begin{figure*}[t]
\centering
\includegraphics[width=1\textwidth]{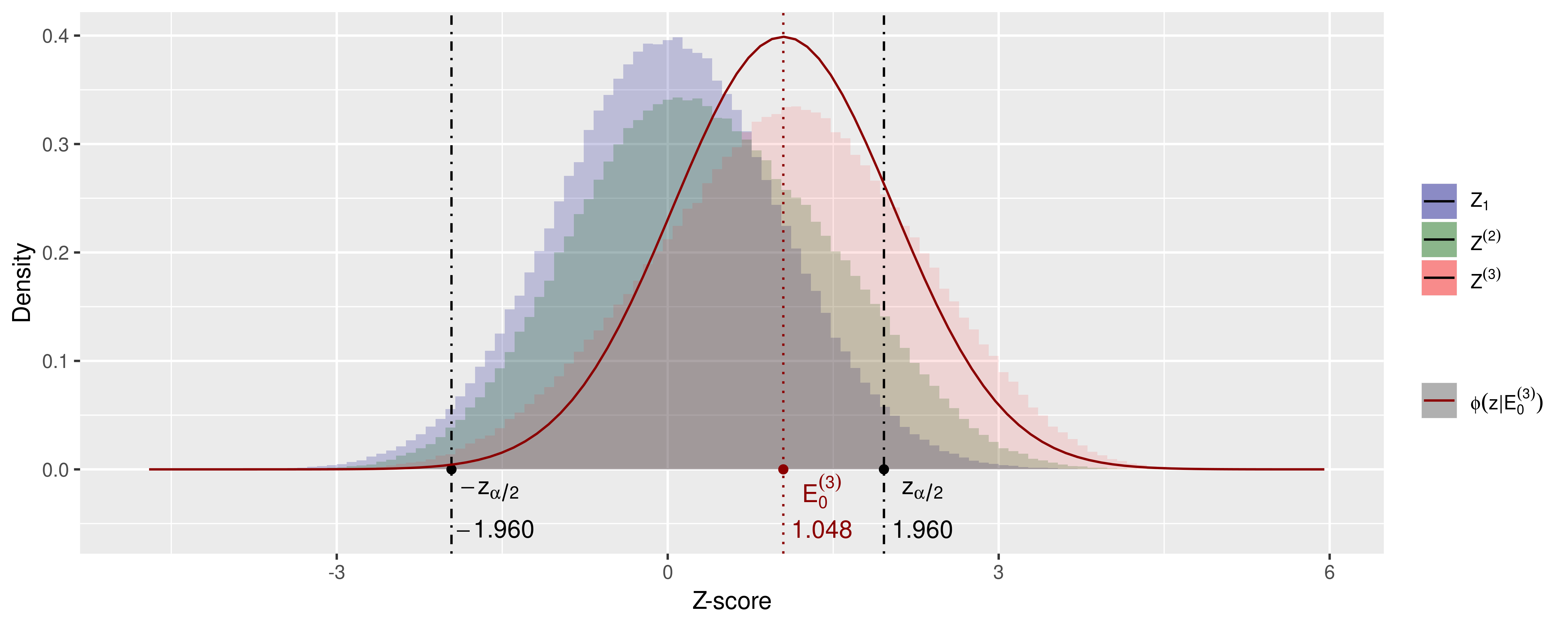}
\caption{\label{fig:distrZ} Sampling distributions of meta-analysis \(Z^{(t)}\)-scores under the null hypothesis in the \lingform{Gold Rush} scenario, under the equal study size assumption, with \(\alpha = 0.05\) and values for \(\omega_{\textsc{s}}^{(1)}\), \(\omega_{\textsc{ns}}^{(1)}\), \(\omega_{\textsc{s}}\) and \(\omega_{\textsc{ns}}\) from Eq. \eqref{newStudyProbs}. \(Z^{(t)}\) is as defined in Eq. \eqref{Zt:Equaln}. \(\phi(z | \ExpNull^{(3)})\) the standard normal density function shifted by \(\ExpNull^{(3)}\), with \(\ExpNull^{(3)}\) shorthand for \(\ExpNull\left[Z^{(3)}\,\middle|\, T \geq 3 \right]\). See Appendix \ref{sec:AppCode} for the code that produces the simulation and creation of this figure.}
\end{figure*}

In the next sections we focus on \lingform{Gold Rush} Accumulation Bias under the null hypothesis and its effect on type-I error control. The values in rightmost column of Eq. \eqref{newStudyProbs} are introduced to obtain estimates for the Accumulation Bias in the test estimates. These values are not supposed to be realistic, but are chosen to demonstrate the effect of Accumulation Bias as clearly as possible. The extreme values 1 for \(\omega_{\textsc{s}}^{(1)}\) and \(\omega_{\textsc{s}}\) given in Eq. \eqref{newStudyProbs} support the simulation of large study series under the null hypothesis. The small values for \(\omega_{\textsc{ns}}^{(1)}\) and \(\omega_{\textsc{ns}}\) are chosen such that the effect of significant findings on the sampling distribution is clearly visible (see Section \ref{sec:GoldDistribution} and Figure \ref{fig:distrZ}). 
For \(\alpha = 0.05\), \(\omega_{\textsc{s}}^{(1)} = 1\) implies that, in expectation under the null distribution, all of the 2.5\% (\(\frac{\alpha}{2}\)) positively significant pilot studies under the null hypothesis become a two-study series, while \(\omega_{\textsc{ns}}^{(1)} = 0.1\) indicates that, since an expected 95\% (\(1-\alpha\)) of pilot studies is not significant under the null hypothesis, \(9.5 \%\) (\(0.1 \cdot 95\%\)) become a two-study series. For study series beyond the pilot study and its replication, this setup entails that in all studies, except for the last and the first, the fraction of significant findings is more than half, since \(\omega_{\textsc{s}} = 0.02\) implies that only \(0.02 \cdot 95\% = 1.9\%\) nonsignificant studies grow into a larger study series: the expected fraction of significant studies in growing series under the null hypothesis converges to \(2.5 / (2.5 + 1.9) = 0.6\).

\begin{table*}[t]
\centering
\caption{\label{tab:ExpZm} Expected \(Z\)-scores under the null hypothesis in the \lingform{Gold Rush} scenario, under the equal study size assumption, calculated using Eq. \eqref{expZt:Equaln} with \(\alpha = 0.05\) and values for \(\omega_{\textsc{s}}^{(1)}\), \(\omega_{\textsc{ns}}^{(1)}\), \(\omega_{\textsc{s}}\) and \(\omega_{\textsc{ns}}\) from Eq. \eqref{newStudyProbs}. \(Z^{(t)}\) is as defined in Eq. \eqref{Zt:Equaln}. See Appendix \ref{sec:AppCode} for the code that was used to calculate these values.}
\begin{filecontents*}{E0Zt.csv}
"","E0.Zt","E0.ZtCond","E0.Zmeta"
1,0,0.487,0
2,0,1.328,0.344
3,0,1.328,1.048
4,0,1.328,1.572
5,0,1.328,2
6,0,1.328,2.368
7,0,1.328,2.695
8,0,1.328,2.99
9,0,1.328,3.262
10,0,1.328,3.515
\end{filecontents*}
\pgfplotstabletypeset[
col sep = comma,
every head row/.style={before row=\toprule, after row=\midrule},
every last row/.style={after row=\bottomrule},
every even row/.style={before row={\rowcolor[gray]{0.9}}},
every head row/.style={before row=\toprule,after row=\midrule},
every last row/.style={after row=\bottomrule},
every head row/.style={before row=\toprule, after row=\midrule},
every last row/.style={after row=\bottomrule},
display columns/0/.style={string type, column name = {Number of\\studies (\(t\))}, column type = {c}},
display columns/1/.style={column name = {\(\ExpNull\left[Z_{t}\right]\)}, column type = {c}},
display columns/2/.style={column name = {\(\ExpNull\left[Z_{t}\,\middle|\, T \geq t + 1 \right]\)}, column type = {c}},
display columns/3/.style={column name = {\(\ExpNull\left[Z^{(t)} \,\middle|\, T \geq t \right]\)}, column type = {c}}
]
{E0Zt.csv}
\end{table*}

\subsection{\lingform{Gold Rush} Accumulation Bias' estimates under the null hypothesis} \label{sec:GoldEstimates}

The new study probability parameters in Eq. \eqref{newStudyProbs} are much larger when results are positively significant than when they are not. As a result, study series that contain more significant studies have larger probabilities to come into existence than those that contain less. While the expectation of a \(Z\)-score is 0 under the null hypothesis for each individual study (for all \(t\): \(\ExpNull\left[Z_{t}\right] = 0\)), the expectation of a study that is part of a series of studies is larger. This shift in expectation introduces the Accumulation Bias in the estimates.

The main ingredient of the bias in the meta-analysis \(Z^{(t)}\)-score is the bias in the individual study \(Z_t\)-scores, conditional on being part of a series. This is already apparent for the pilot study, which we use as an example by expressing its expected value under the null hypothesis, given that it has a successor study: \(\ExpNull\left[Z_1 \,\middle|\, T \geq 2\right]\). This conditional expectation is a weighted average of two other expectations that are conditioned further based on the events that lead to a new study according to Eq. \eqref{newStudyProbs}: \(\ExpNull\left[Z_1\,\middle|\, Z_1 \geq z_{\frac{\alpha}{2}}\right]\), \(Z_1\) from the right tail of the null distribution, and the nonsignificant results with expectation \(\ExpNull\left[Z_1 \,\middle|\, \abs{Z_1} < z_{\frac{\alpha}{2}} \right]\). We discard negative significant results, since those were given 0 probability to produce replication studies in Eq. \eqref{newStudyProbs}. The positive significant and nonsignificant results are weighted by the new study probabilities in Eq. \eqref{newStudyProbs} and the probabilities under the null distribution of sampling from either the tail (\(\alpha\)) or the middle part (\(1-\alpha\)) of the standard normal distribution. A more detailed specification of these components can be found in Appendix \ref{sec:AppExpPilot}. If we assume a significance threshold of 5\% we obtain:
\begin{equation} \label{Z1}
\begin{split}
& \text{For } \alpha = 0.05: \\
& \ExpNull\left[Z_1 \,\middle|\, T \geq 2\right] \\
& \quad = \frac{\int_{z_{\frac{\alpha}{2}}}^{\infty}{z \cdot \phi(z)dz} \cdot \omega_{\textsc{s}}^{(1)} \cdot \frac{\alpha}{2} + 0 \cdot \omega_{\textsc{ns}}^{(1)} \cdot (1-\alpha)}{\omega_{\textsc{s}}^{(1)} \cdot \frac{\alpha}{2} + \omega_{\textsc{ns}}^{(1)} \cdot (1-\alpha)} \approx 0.487.
\end{split}
\end{equation}
Here we use the fact that, for \(\alpha = 0.05\), \(\ExpNull\left[Z_1 \,\middle|\, Z_1 \geq z_{\frac{\alpha}{2}}\right] = \int_{1.960}^{\infty}z \cdot \phi(z)dz \approx 2.338\), with \(\phi()\) the standard normal density function and that \(\ExpNull\left[Z_1 \,\middle|\, \abs{Z_1} < z_{\frac{\alpha}{2}} \right]\) is the expectation of a symmetrically truncated standard normal distribution, which is 0. The value 0.487 is obtained by using the parameter values given in Eq. \eqref{newStudyProbs}. For studies in the series later than the pilot study, the expression follows analogously by taking for all \(t \geq 2:\) \(\omega_{\textsc{s}}^{(t)} = \omega_{\textsc{s}} \text{ and } \omega_{\textsc{ns}}^{(t)} = \omega_{\textsc{ns}}\): \(\ExpNull\left[Z_{t} \,\middle|\, T \geq t+1 \right] \approx 1.328\).

To determine the effect on the meta-analysis \(Z^{(t)}\)-score, we define the expectation under the null hypothesis \(\ExpNull\left[Z^{(t)} \,\middle|\, T \geq t \right]\), conditioned on the availability of a series of size \(t\). To specify this expectation, we use that the last study is always unbiased since we do not know whether it will spur more studies. As shown in more detail in Appendix \ref{sec:AppExpMeta}, the expression follows from Eq. \eqref{Zt:Unequaln} by separately treating the unbiased expectation of 0 and the pilot study. If we assume a significance threshold of 5\%, we obtain the general expression in Eq. \eqref{expZt:Unequaln} and the expression in Eq. \eqref{expZt:Equaln} under the assumption of equal study sizes (\(n_1 = n_2 = \dots = n_t = n\)):

\begin{subequations}
\begin{align}
\nonumber
& \text{For } \alpha = 0.05, \text{ for all } t \geq 2: \\
\nonumber
& \ExpNull\left[Z^{(t)}\,\middle|\, T \geq t \right] \\
\label{expZt:Unequaln}
& \quad \approx 
\frac{\sqrt{n_{1}} \cdot 0.487 + 
\sum_{i = 2}^{t-1} \sqrt{n_{i}} \cdot 1.328 + \sqrt{n_{t}} \cdot 0}{\sqrt{N^{(t)}}} \\
\label{expZt:Equaln}
& \quad = \frac{0.487 + 1.328(t - 2)}{\sqrt{t}}.
\end{align}
\end{subequations}
Table \ref{tab:ExpZm} shows the Accumulation Bias in the estimates of \(\ExpNull\left[Z^{(t)}\,\middle|\, T \geq t \right]\) as studies accumulate under the \lingform{Gold Rush} scenario, with equal study sizes and values for the new study probabilities given by Eq. \eqref{newStudyProbs}.

\subsection{\lingform{Gold Rush} Accumulation Bias' sampling distribution under the null hypothesis} \label{sec:GoldDistribution}

Figure \ref{fig:distrZ} shows simulated \lingform{Gold Rush} sampling distributions for study series of size two and three in comparison to an individual study \(Z\)-distribution. Because the new study probabilities in Eq. \eqref{newStudyProbs} give \(Z_{t-1}\)-values below \(\minus z_{\frac{\alpha}{2}}\) zero probability to warrant a successor study, values for the \(z^{(t)}\)-statistic below \(\minus z_{\frac{\alpha}{2}}\) will be scarce and the larger \(t\) is the larger this scarcity will be since only the last study is able to provide such small \(Z\)-score estimates. The opposite is the case for values above \(z_{\frac{\alpha}{2}}\), which have probability 1 to warrant a new study. As a result, the distribution of the meta-analysis \(Z\)-score has negative skew (more mass on the right, more tail to the left). See the comparison to the normal distribution also plotted in Figure \ref{fig:distrZ} for a three-study series. Skewness is not the only characteristic that distinguishes the resulting distribution from a standard normal. The variance also deviates since the meta-analysis distribution is a mixture distribution.

\begin{table*} [t]
\centering
\caption{\label{tab:TypeIerrorTildeSim} Inflated type-I error rates for tests affected by bias only and tests affected by bias as well as impaired sampling distribution. Simulated values are under the null hypothesis in the \lingform{Gold Rush} scenario, under the equal study size assumption, with \(\alpha = 0.05\) and values for \(\omega_{\textsc{s}}^{(1)}\), \(\omega_{\textsc{ns}}^{(1)}\), \(\omega_{\textsc{s}}\) and \(\omega_{\textsc{ns}}\) from Eq. \eqref{newStudyProbs}. See Appendix \ref{sec:AppCode} for the code that produces the simulation and creation of this table.}
\begin{filecontents*}{TypeIerror.csv}
"","TypeIerrorTilde","TypeIerrorSim"
2,0.06,0.1
3,0.18,0.23
4,0.35,0.4
5,0.52,0.53
\end{filecontents*}
\pgfplotstabletypeset[
precision = 2,
col sep = comma,
every head row/.style={before row=\toprule, after row=\midrule},
every last row/.style={after row=\bottomrule},
every even row/.style={before row={\rowcolor[gray]{0.9}}},
every head row/.style={before row=\toprule,after row=\midrule},
every last row/.style={after row=\bottomrule},
every head row/.style={before row=\toprule, after row=\midrule},
every last row/.style={after row=\bottomrule},
display columns/0/.style={string type, column name = {Number of studies (\(t\))}, column type = {c}},
display columns/1/.style={column name = {\(\widetilde{\ProbNull}[\cE^{(t)}_{\textsc{type-I}} \mid T \geq t ]\)}, column type = {c}},
display columns/2/.style={column name = {\(\ProbNull[\cE^{(t)}_{\textsc{type-I}} \mid T \geq t ]\)}, column type = {c}}
]
{TypeIerror.csv}
\end{table*}
For a two-study meta-analysis \(Z^{(2)}\) we obtain a mixture of two conditional distributions, one conditioned on the first study being a significant --- sampled from the right tail of the distribution (with probability \(\frac{\alpha}{2} \cdot \omega_{\textsc{s}}^{(1)}\)) --- and one with the first study nonsignificant --- sampled from the symmetrically truncated normal distribution (with probability \((1 - \alpha) \cdot \omega_{\textsc{ns}}^{(1)}\)). Because the combined distribution on \(Z^{(2)}\) is a mixture of the two scenarios, its variance is larger than the variance of either of the two components of the mixture, as we show in Appendix \ref{sec:AppMixVar}. In Figure \ref{fig:distrZ} we see that, with the parameter values from Eq. \eqref{newStudyProbs} the variance of \(Z^{(2)}\) and \(Z^{(3)}\) are even larger than that of \(Z_1\), even though both \(\Var\left\{Z^{(2)} \,\middle|\, Z_1 < z_{\frac{\alpha}{2}} \right\}\) and \(\Var\left\{Z^{(2)} \,\middle|\, \abs{Z_1} \geq z_{\frac{\alpha}{2}} \right\}\) are smaller. Hence the sampling distribution under the null hypothesis of a meta-analysis \(Z\)-score deviates from a standard normal under Accumulation Bias due to a non-zero location (the bias), skewness and inflated variance. All three inflate the probability of a type-I error in a standard normal test, as we will study in the next section.

\subsection{\lingform{Gold Rush} Accumulation Bias' influence on p-value tests} \label{sec:GoldTesting}

Let us now establish the effect of our \lingform{Gold Rush} Accumulation Bias on meta-analysis testing when using common/fixed-effects \(Z\)-tests. Let \(\cE^{(t)}_{\textsc{type-I}}\) indicate the event of a type-I error (significant result under the null hypothesis) in a meta-analysis of \(t\) studies and let \(\ProbNull\left[\cE^{(t)}_{\textsc{type-I}} \,\middle|\, T \geq t \right] = \ProbNull\left[\abs{Z^{(t)}} \geq z_{\frac{\alpha}{2}} \,\middle|\, T \geq t \right]\) denote the expected rate of type-I errors in a two-sided common/fixed-effect \(Z\)-test for studies \(i\) up to \(t\) conditional on the fact that at least \(t\) studies were performed. 

We obtain the type-I error rate for this test by simulating the \lingform{Gold Rush} scenario, for which the results are shown in the right hand column of Table \ref{tab:TypeIerrorTildeSim}, assuming \(\alpha = 0.05\). If only bias would be at play, the sampling distribution under the null hypothesis would be a shifted normal distribution. Eq. \eqref{TypeIerrorBiasCorrected} expresses the expected type-I error rate for this bias only scenario, with \(\Phi()\) the cumulative normal distribution. The inflation actual inflation in the type-I error rate is larger than shown by this scenario, as illustrated the Table \ref{tab:TypeIerrorTildeSim}. The difference between these two type-I error rates for a series of three studies is depicted in Figure \ref{fig:distrZ} by the area under the red histogram for \(Z^{(3)}\) and the red \(\phi(z \mid \ExpNull^{(3)})\) curve below \(\minus z_{\frac{\alpha}{2}}\) and above \(z_{\frac{\alpha}{2}}\). We conclude that the effect of Accumulation Bias on testing cannot be corrected by only an approximation of the bias.

\begin{equation} \label{TypeIerrorBiasCorrected}
\begin{split}
\widetilde{\ProbNull}\left[\cE^{(t)}_{\textsc{type-I}} \,\middle|\, T \geq t \right] &:= 1 - \Phi\left(z_{\frac{\alpha}{2}} - \ExpNull\left[Z^{(t)}\,\middle|\, T \geq t \right]\right) \\
 & \quad + \Phi\left(\minus z_{\frac{\alpha}{2}} - \ExpNull\left[Z^{(t)}\,\middle|\, T \geq t \right]\right).
\end{split}
\end{equation}

\subsection{\lingform{Gold Rush} Accumulation Bias: When does it occur?} \label{sec:GoldOccur}
We indicated in Section \ref{sec:GoldNewStudyProbsInd} that we chose extreme values for parameters \(\omega_{\textsc{s}}^{(1)}\), \(\omega_{\textsc{x}}^{(1)}\), \(\omega_{\textsc{ns}}^{(1)}\), \(\omega_{\textsc{s}}\), \(\omega_{\textsc{x}}\) and \(\omega_{\textsc{ns}}\) such that Figure \ref{fig:distrZ} would clearly show the bias and distributional change that occurs. However, for any combination of values for which there is a \(t\) where \(\omega_{\textsc{s}}^{(t)} \neq \omega_{\textsc{x}}^{(t)} \neq \omega_{\textsc{ns}}^{(t)}\) Accumulation Bias occurs for series larger than size \(t\) and p-value tests that assume a standard normal distribution are invalid.

\section{The Accumulation Bias Framework} \label{sec:Framework}

In general, Accumulation Bias in meta-analysis makes the sampling distribution of the meta-analysis \(Z\)-score difficult to characterize due to the data dependent size and timing of a study series up for meta-analysis. In this section, we specify both processes in a framework of analysis time probabilities. We use the term \lingform{analysis time} because time in meta-analysis is partly based on a \lingform{survival time}. A survival time indicates that a subject lives longer than time \(t\) (and might still become much older), just as an analysis time indicates that a series up for meta-analysis has at least size \(t\) (but might still grow much larger). As such, analysis time probabilities, just as the probabilities in a survival function, do not add up to 1.

Our \lingform{Accumulation Bias Framework} uses the following notation for its three key components: \(S(t-1)\), \(\cA^{(t)}\) and \(A(t)\). Firstly, \(S(t-1)\) can be understood as the survival function in the variable time \(t\) that indicates the size of the expanding study series. \(S(t-1)\) denotes the probability that the available number of studies is at least \(t\) (\(\Prob[T \geq t]\)), so the study series has survived past the previous study at \(t - 1\). Secondly, \(\cA^{(t)}\) indicates the event that a meta-analysis is performed on a study series of size exactly \(t\). Lastly, \(A(t)\) combines the probability that a study series of certain size is available (\(S(t-1)\)) with the decision \(\cA^{(t)}\) to perform the analysis on exactly \(t\) studies. So the \lingform{analysis time probability} \(A(t)\) represents the general probability that a meta-analysis of size \(t\) --- so at \lingform{time} \(t\) --- is performed and is the key to describing the influence of various forms of Accumulation Bias on testing.

\subsection{Analysis time probabilities}

Let \(\Prob\left[\cA^{(t)} \,\middle|\, T \geq t, z_1, \ldots, z_t \right]\) denote the probability that a meta-analysis is performed on the first \(t\) studies. Just as the \lingform{Gold Rush}' new study probabilities from Eq. \eqref{newStudyProbs}, this probability can depend on the results in the study series \(z_1, \ldots, z_t \). The event \(\cA^{(t)}\) only occurs if a series of size \(t\) is available, so we need to condition on the survival past \(t - 1\), which can also depend on previous results. When combined, we obtain the following definition\footnote{Note that $A(t \mid z_1, \ldots, z_t )$ is defined as a product of two (conditional) probabilities. Calling this product itself a \enquote{probability}, as we do, can be justified as follows: we currently think of the decision whether to continue studies at time $t$, i.e. whether $T \geq t$, to be made before the $t$-th study is performed. But we may also think of the $t$-study result $z_t$ as being generated irrespective of whether $T \geq t$, but remaining unobserved for ever if $T < t$. If the decision whether $T \geq t$ is made independently of the value $z_t$, i.e. we add the constraint $\Prob\left[ T \geq t \,\middle|\, z_1, \ldots, z_{t-1} \right]=\Prob\left[ T \geq t \,\middle|\, z_1, \ldots, z_{t} \right]$, then the resulting model is mathematically equivalent to ours (in the sense that we obtain exactly the same expressions for $S(t)$, $A(t \mid z_1,\ldots, z_t)$, all error probabilities etc.), but it does allow us to write, by Eq. \eqref{eq:anaTimeProb}, that $A\left(t \,\middle|\, z_1, \ldots, z_t \right)= \Prob \left[\cA^{(t)} , T \geq t \,\middle|\, z_1, \ldots, z_t \right]$ --- so now $A\left(t \,\middle|\, z_1, \ldots, z_t \right)$ is indeed a probability.} of \lingform{analysis time probabilities} \(A(t)\):
\begin{equation}\label{eq:anaTimeProb}
\begin{split}
A\left(t \,\middle|\, z_1, \ldots, z_t \right) &:= \Prob\left[\cA^{(t)} \,\middle|\, T \geq t, z_1, \ldots, z_t \right] \\
& \quad\cdot 
S\left(t - 1 \,\middle|\, z_1, \ldots, z_{t-1}\right), \\ \\
\quad \text{where we define } \\
S\left(t - 1 \,\middle|\, z_1, \ldots, z_{t-1}\right) &:= \Prob\left[ T \geq t \,\middle|\, z_1, \ldots, z_{t-1} \right].
\end{split}
\end{equation}
Eq. \eqref{eq:anaTimeProb} formalizes the idea of analysis time probabilities ``depending on previous results'' in terms of the individual study \(Z\)-scores \(z_1, \ldots, z_t \). This is compatible with the \(Z\)-test approach in meta-analysis and the dependencies and the \lingform{Gold Rush}' new study probabilities that are explicitly expressed in terms of \(Z\)-scores. More generally however, in Section \ref{sec:Size} and \ref{sec:Timing} we extend the definition and allow analysis time probabilities to also depend on the data in the original scale and external parameters.

\subsection{Analysis time probabilities' independence from the data-generating hypothesis} \label{sec:AnaTimeInd}
Just as for the \lingform{Gold Rush}' new study probabilities discussed in Section \ref{sec:GoldNewStudyProbs} and \ref{sec:GoldNewStudyProbsInd}, the analysis time probabilities \(A(t)\) only depend on the data, and are independent from the hypothesis that generated the data. So again, \(\Prob\) in these definitions can be read as \(\ProbAlt\) as well as \(\ProbNull\). Our definition of \(A(t)\) relates to the definition of a \lingform{Stopping Rule} by \citet[pp. 33-34]{berger1988relevance}, where they use \(x^{(m)}\) to denote a vector of \(m\) observations:
\begin{quote}\textbf{Definition.} A \lingform{stopping rule} is a sequence \(\tau = \left(\tau_{0}, \tau_{1}, \dots\right)\) in which \(\tau_{0} \in \left[0, 1\right]\) is a constant and \(\tau_{m}\) is a measurable function of \(x^{(m)}\) for \(m \geq 1\), taking values in \(\left[0, 1\right]\).
\end{quote}
\begin{quote}
\(\tau_0\) is the probability of stopping the experiment with no observations (e.g., if it is determined that the experiment is too expensive); \(\tau_1(x^{(1)})\) is the probability of stopping after observing the datum \(x^{(1)} = x_1\), conditional on having taken the first observation; \(\tau_2(x^{(2)})\) is the probability of stopping after observing \(x^{(2)} = (x_1, x_2)\), conditional on having taken the first and second observations; etc.
\end{quote}
To take the analogy with survival analysis further, we consider the sequence \(\tau\) defined above by \citet{berger1988relevance} to be a sequence of hazards. Instead of using their notation \(\tau\) we denote the \lingform{Stopping Rule} by \(\lambda = \left(\lambda(0), \lambda(1), \dots\right)\) to emphasize its behavior as a sequence of \lingform{hazard functions} and to distinguish time \(t\) from the probability \(\lambda(t)\) of stopping at that time given that you were able to reach it. The hazard of stopping at time \(t\) can depend on previous results and is defined as follows:

\begin{equation}
    \lambda\left(t \,\middle|\, z_1, \ldots, z_t \right) := \Prob\left[T = t \,\middle|\, T \geq t, z_1, \ldots, z_t \right].
\end{equation}
In this paper we are only interested in cases in which a first study is available, so \(\lambda(0) = 0\) (also stated as \(\Prob[T \geq 1] = 1\) in Appendix \ref{sec:AppExpPilot}). The survival \(S(t - 1)\), the probability of obtaining a series of size at least \(t\) (so larger than \(t - 1\)), follows from the hazards by considering that surviving past time \(t - 1\) means that the series has not stopped at studies \(i\) up to and including \(t - 1\). So for \(t \geq 1\):

\begin{equation} \label{hazard}
    S\left(t - 1 \,\middle|\, z_1, \ldots, z_{t-1}\right) = \prod_{i = 0}^{t - 1} (1 - \lambda\left(i \,\middle|\, z_1, \ldots, z_i\right)).
\end{equation}
In many examples, the hazard of stopping at time \(t\), \(\lambda(t)\), will depend on the result \(z_t\) just obtained. In that case \(\lambda\left(i \,\middle|\, z_1, \ldots, z_i\right) = \lambda\left(i \,\middle|\, z_i\right)\) in Eq. \eqref{hazard} above. But in general \(\lambda(t)\) might also depend on some synthesis of all \(z_i\) so far. We show some of the variety of forms that \(\lambda(t)\), \(S(t)\) and \(A(t)\) can take in our Accumulation Bias Framework in the following sections.

\subsection{Accumulation Bias caused by dependent study series size} \label{sec:Size}
Our \lingform{Gold Rush} example describes an instance of Accumulation Bias that is caused by how the study series size comes about. This is expressed by the \(S(t)\) component of the analysis times probability \(A(t)\). We represent our \lingform{Gold Rush} scenario in terms of our Accumulation Bias framework in next section, followed by variations from the literature that we were able to express in a similar manner.

\subsubsection{\lingform{Gold Rush}: dependence on significant study results \label{sec:GoldAnaTimes}}

The \lingform{Gold Rush} scenario operates in a general meta-analysis setting and assumes that there is a single random or prespecified time \(t\) at which a study series is up for meta-analysis. This is the approach taken by meta-analyses not explicitly part of a living systematic review. In the \lingform{Gold Rush} example the dependency arises in the study series because a \(t\)-study series has a larger probability to come into existence when individual study results are significant, and you need a \(t\)-study series to perform a \(t\)-study meta-analysis. This dependency was characterized by the new study probabilities \(\omega_{\textsc{s}}^{(1)}\), \(\omega_{\textsc{ns}}^{(1)}\), \(\omega_{\textsc{s}}\) and \(\omega_{\textsc{ns}}\) from Eq. \eqref{newStudyProbs}. The value of \(S(t)\), and therefore \(A(t)\), can be expressed in terms of these new study probabilities by considering whether \(z_1, \ldots, z_{t-1}\) are larger than \(z_{\frac{\alpha}{2}}\) (which is 1.960 for \(\alpha\) = 0.05). Since a meta-analysis is performed only once at a randomly chosen time \(t\), we have \(\Prob[\cA^{(t)}] = 1\) for that time \(t\) and \(\Prob[\cA^{(t)}] = 0\) otherwise. So for the one meta-analysis we obtain:

\begin{equation}
\begin{split}
\text{For \(t\) such that } \Prob[\cA^{(t)}] = 1:& \\
A\left(t \,\middle|\, z_1, \ldots, z_{t-1}; \alpha\right) =& S\left( t - 1 \,\middle|\, z_1, \ldots, z_{t-1}; \alpha\right) \\
=& \prod_{i = 0}^{t - 1} \left(1 - \lambda\left(i \,\middle|\, z_i;\alpha\right)\right),
\end{split}
\end{equation}
with \(\lambda\left(0\right) = 0\) and for all \(i \geq 1\), \(\lambda(i)\) is defined as follows:
\begin{equation}
\begin{split}
\lambda\left(i \,\middle|\, z_i, \alpha\right) =& \ 1 - \left( \omega^{(i)}_{\textsc{s}} \cdot \mathbbm{1}_{z_i \geq z_{\frac{\alpha}{2}}} + \omega^{(i)}_{\textsc{ns}} \cdot \mathbbm{1}_{\abs{z_i} < z_{\frac{\alpha}{2}}} \right) \\
\overline{\lambda}_0 \left(i \,\middle|\, \alpha\right) :=& \ \ExpNull\left[\lambda(i \,\middle|\, Z_i; \alpha)\right] \\
=& \ 1 - \left( \omega^{(i)}_{\textsc{s}} \cdot \frac{\alpha}{2} + \omega^{(i)}_{\textsc{ns}} \cdot (1-\alpha) \right).
\end{split}
\end{equation}
Therefore, (leaving out the \(\lambda(0)\) and summing from \(i = 1\) to \(t-1\)), we obtain the following expressions for the \lingform{Gold Rush} analysis time probabilities and its expectations under the null distribution:
\begin{equation} \label{A(t)Gold Rush}
\begin{split}
A\left(t \,\middle|\, z_1, \ldots, z_{t-1}; \alpha\right) =& 
\prod_{i = 1}^{t - 1} \left( \omega_{\textsc{s}}^{(i)} \cdot \mathbbm{1}_{z_i \geq z_{\frac{\alpha}{2}}} + \omega_{\textsc{ns}}^{(i)} \cdot \mathbbm{1}_{\abs{z_i} < z_{\frac{\alpha}{2}}} \right)\\ \\
\overline{A}_0\left(t \,\middle|\, \alpha\right) :=& \ExpNull\left[A(t \mid Z_1, \ldots, Z_{t-1}; \alpha)\right] \\
=& \prod_{i=1}^{t-1} 
\left(\omega_{\textsc{s}}^{(i)} \cdot \frac{\alpha}{2} + \omega^{(i)}_{\textsc{ns}} \cdot (1-\alpha)\right).
\end{split}
\end{equation}

\subsubsection{\citet{kulinskaya2016sequential}: dependence on meta-analysis estimates} \label{sec:Kulinskaya}

\citet{kulinskaya2016sequential} report biases that result from dependencies between a current meta-analysis estimate and the decision to perform a new study. Since their focus is on bias, they do not discuss issues of multiple testing over time, which would arise if their cumulative meta-analyses estimates were tested. In this section we assume that the timing of the meta-analysis test is independent from the estimates that determined the size of the series, as if a test were done by a second unknowing meta-analyst. This scenario is hinted at by \citet[p. 296]{kulinskaya2016sequential} in the statement \enquote{When a practitioner or a meta-analyst finds several trials in the literature, a particular decision-making scenario may have already taken place.} We postpone the discussion of multiple testing to Section \ref{sec:LSR}. In this estimation setting, the decision to perform new studies is determined not by the meta-analysis \(Z\)-scores \(Z^{(t-1)}\), but by the meta-analysis estimates on the original scale \(M^{(t - 1)}\) (notation adopted from \citet{borenstein2009introduction}, see Appendix \ref{sec:AppCommFixMetaAnalysis}), in relation to a minimally clinically relevant effect \(\Delta^{H1}\). A minimally clinically relevant effect is the effect that should be used to power a trial (in the alternative distribution H1), and therefore, the effect that the researchers of the study do not want to miss. \citet{kulinskaya2016sequential} consider three models for the study series accumulation process: the \lingform{power-law model} and the \lingform{extreme-value model} and the \lingform{probit model}. The models relate the probability of a new study to the cumulative meta-analysis estimate of the study series so far and are inspired by models for publication bias. Although all three models can be recast in our framework, we demonstrate this only for the power law model that uses one extra parameter \(\tau\) to relate the previous meta-analysis estimate \(M_{(t-1)}\) to \(S(t)\). Just as in the \lingform{Gold Rush} scenario, we must assume that a meta-analysis test is performed only once at a randomly chosen time \(t\). So only at that time \(t\) \(\Prob[\cA^{(t)}] = 1\) and \(\Prob[\cA^{(t)}] = 0\) otherwise. We obtain the following expression for the \citet{kulinskaya2016sequential} \lingform{power-law model}:

\begin{equation}\label{kulinskaya}
\begin{split}
\text{For \(t\) such that }\Prob[\cA^{(t)}] =& 1: \\
A\left(t \,\middle|\, M^{(t - 1)}; \Delta^{H1}, \tau \right) =& \ S\left(t - 1 \,\middle|\, M^{(t - 1)}; \Delta^{H1}, \tau \right) \\
=& \prod_{i = 0}^{t - 1} (1 - \lambda\left(i \,\middle|\, M^{(t - 1)}; \Delta^{H1}, \tau\right)),
\end{split}
\end{equation}
with \(\lambda(0) = \lambda(1) = 0\), and for all \(i \geq 2\), \(\lambda(i)\) is defined as follows:
\begin{equation}
\begin{split}
\quad \lambda\left(i \,\middle|\, M^{(i - 1)}; \Delta^{H1}, \tau\right)& = 1 - \left(\frac{M^{(i - 1)}}{\Delta^{H1}}\right)^{\tau},
\end{split}
\end{equation}
for \(0 < M^{(i-1)} < \Delta^{H1}\) and 1 (so \(1 - \lambda = 0\)) otherwise.

According to this model, no further studies are performed as soon as an estimate as large as \(\Delta^{H1}\) is found. For estimates smaller than \(\Delta^{H1}\), the closer the estimate is to \(\Delta^{H1}\), the larger the probability of a subsequent study. Just as in the \lingform{Gold Rush} example, this model will introduce bias as well as skew the sampling distribution of the data under the null hypothesis since initial studies with large estimates have larger probability to end up in study series of considerable size than small initial estimates do. When the initial study gives a large overestimation of the effect, this overestimation stays present in the subsequent meta-analysis estimates and keeps influencing the probability of subsequent studies. Therefore, this model shows the effect of early studies in the series even more clearly than the \lingform{Gold Rush} example. However, the accumulation bias does have a cap, since estimates larger than \(\Delta^{H1}\) do not introduce new replication studies.

\subsubsection{\citet{whitehead2002meta}: dependence on early study results \label{sec:Whitehead}}
\begin{quote}
   Bias may also be introduced by the order in which studies are conducted. For example, large-scale clinical trials for a new treatment are often undertaken following promising results from small trials. [...] given that a meta-analysis is being undertaken, larger estimates of treatment difference are more likely from the small early studies than from the later larger studies.
   --Whitehead (\citeyear[p. 197]{whitehead2002meta})
\end{quote}
\citet{whitehead2002meta} mentions a dependence between the results of the small early studies in a series and the size of the series. This influence could either be based on the significance of early findings, such as in the \lingform{Gold Rush} example (Section \ref{sec:GoldAnaTimes}), or on the estimates in the initial studies, such as in the power law model from \cite{kulinskaya2016sequential} (Section \ref{sec:Kulinskaya}). \cite{whitehead2002meta} does not give sufficient details to specify this dependency explicitly, but we are confident that it will fit in our Accumulation Bias framework.

Two ways to approach this Accumulation Bias are given in \cite{whitehead2002meta}. The first is to exclude early studies from the meta-analyses, either in the main analysis or in a sensitivity analysis. The second way is to ignore the problem, since the small studies will have little effect on the overall estimate. In Section \ref{sec:LR} we show that any small initial study dependency that can be expressed in terms of \(A(t)\) can be dealt with by tests using likelihood ratios.

\subsubsection{Living Systematic Reviews: dependence on significant meta-analyses + multiple testing} \label{sec:LSR}
\begin{quote}
A living systematic review (LSR) should keep the review current as new research evidence emerges. Any meta-analyses included in the review will also need updating as new material is identified. If the aim of the review is solely to present the best current evidence standard meta-analysis may be sufficient, provided reviewers are aware that results may change at later updates. If the review is used in a decision-making context, more caution may be needed. When using standard meta-analysis methods, the chance of incorrectly concluding that any updated meta-analysis is statistically significant when there is no effect (the type I error) increases rapidly as more updates are performed. --Simmonds, Salanti, McKenzie \& Elliott (\citeyear[p. 39]{simmonds2017living})
\end{quote}
In living systematic reviews, the aim is to have a meta-analysis available to present the current evidence, thus synthesizing the \(t\) studies available at a certain time. The current meta-analysis estimate might be used to decide whether further studies should be performed. In that case \(S(t - 1)\), the probability that a study series of size \(t\) is available --- so that a study series has expanded beyond series size \(t - 1\) --- depends on the meta-analysis estimate \(Z^{(t - 1)}\) at the previous study's meta-analysis. Because the review is continuously updated, \(\Prob[\cA]\) is always 1, and living systematic reviews can be described by the following analysis time probability \(A(t)\):

\begin{equation} \label{LSR}
\begin{split}
A\left(t \,\middle|\, z^{(1)}, \ldots, z^{(t)}; z_{\frac{\alpha}{2}} \right) &= \Prob\left[\cA^{(t)} \,\middle|\, T \geq t \right] \\
& \qquad \cdot S\left( t - 1 \,\middle|\, z^{(1)}, \ldots, z^{(t)}; z_{\frac{\alpha}{2}} \right) \\
&= S\left( t - 1 \,\middle|\, z^{(1)}, \ldots, z^{(t-1)}; z_{\frac{\alpha}{2}} \right) \\
&= \prod_{i = 0}^{t - 1} (1 - \lambda\left(i \,\middle|\, z^{(i)}; z_{\frac{\alpha}{2}} \right)). 
\end{split}
\end{equation}
The quote above warns against decisions based on the continuously updated meta-analysis using a fixed threshold \(z_{\frac{\alpha}{2}}\). Living systematic reviews experience multiple testing problems of a kind that are familiar from statistical monitoring of individual clinical trials \citep{proschan2006statistical}. If the study series is stopped as soon as a significance threshold is reached, and the obtained meta-analysis is considered the final one, then this final meta-analysis test has an increased chance of a type-I error. So the warning is not to use the following simple stopping rule:

\begin{equation} \label{LSRinflated}
\begin{split}
\quad \lambda\left(i \,\middle|\, z^{(i)}; z_{\frac{\alpha}{2}} \right) &= \mathbbm{1}_{\abs{Z^{(i)}} \geq z_{\frac{\alpha}{2}}}.
\end{split}
\end{equation}
Various corrections to significance thresholds are proposed that relate intermediate looks to a maximum sample size or information size. These corrected thresholds depend on \(\alpha\) and the fraction of sample size or information size available at time \(t\). Examples of such methods are \lingform{Trial sequential analysis} \citep{brok2008apparently, thorlund2008can, wetterslev2008trial} and \lingform{Sequential meta-analysis} \citep[Ch. 12]{whitehead2002meta} \citep{whitehead1997prospectively, higgins2011sequential}. For an overview see \citet{simmonds2017living}. In general, Eq. \eqref{LSR} and \eqref{LSRinflated} show that any dependency between \enquote{the best current evidence} and the accumulation of future studies is part of our Accumulation Bias Framework. We discuss the approach to error control taken by the corrected thresholds in Section \ref{sec:Surv}.

\subsection{Accumulation Bias caused by dependent meta-analysis timing} \label{sec:Timing}
We described various forms of Accumulation Bias that are caused by how the study series size comes about, but dependencies are also introduced by how the meta-analysis itself arises. This is expressed by the \(\Prob\left[\cA^{(t)}\right]\) component of the analysis times probabilities \(A(t)\). We only found one such process mentioned in the literature and will discuss it in the next section.

\subsubsection{\cite{ellis2009temporal}: dependence on the right amount of positive findings}
\begin{quote}
Meta-analysis times are subtle. A train of negative findings would generally not stimulate a meta-analysis. Nor would a string of very positive findings. [...] All this makes the analysis of explicitly defined meta-analysis times very difficult. We conclude that study of bias in meta-analysis based on parametric modeling of meta-analysis times is problematical.
--Ellis \& Stewart (\citeyear[pp. 2454-2455]{ellis2009temporal})
\end{quote}
\citet{ellis2009temporal} do not give an explicit model that we can interpret in terms of \(A(t)\), but indicate that it should depend on the study findings \(Z_i\), or in the original scale, \(\overline{D}_i\) (notation adapted from \citet{borenstein2009introduction}, see Appendix \ref{sec:AppCommFixMetaAnalysis}). Given the quote above, the amount of very positive findings should not be too large, and not too small. Though exact parametric modeling indeed stays problematical, we can assume that a positive finding is a study estimate larger than the minimally clinically relevant effect \(\Delta^{H1}\), define the right amount of positive findings to be in the region [a, b], and show that this fits in our Accumulation Bias Framework by expressing a possible model for \(A(t)\):
\begin{equation}
\begin{split}
\text{For } t \text{ such that } S(t-1) =& 1: \\
A\left(t \,\middle|\, \overline{D}_1, \ldots, \overline{D}_t; a,b\right) =& \Prob\left[\cA^{(t)} \,\middle|\, T \geq t, \overline{D}_1, \ldots, \overline{D}_t; a,b\right] \\
& \quad \cdot S\left( t - 1 \,\middle|\, \overline{D}_1, \ldots, \overline{D}_{t-1}; a, b\right) \\
=& \Prob\left[\cA^{(t)} \,\middle|\, T \geq t, \overline{D}_1, \ldots, \overline{D}_t; a,b\right] \\
=& \mathbbm{1}_{C \in \, [a, b]} \\
&\text{with} \quad C =\sum_{i = 1}^t \mathbbm{1}_{\overline{D}_i > \Delta^{H1}}.
\end{split}
\end{equation}

\begin{table*}[t]
\centering
\caption{\label{tab:Vis} Possible 2001 state of a database of study series per topic, visualizing what study series are taken into account in the two approaches to error control: conditional on time (blue and grey) and surviving over time (orange).}
\begin{tabular}{llllllllllllllll}
                                                                             &                                               & \multicolumn{14}{c}{\textbf{Topics}}                                                                                                                                                                                                                                                                                                                                                                                                                                                                                                                                                                                    \\
                                                                             & \textbf{}                                     & \textbf{1}                                                & \textbf{2}                           & \textbf{3}                           & \textbf{4}                                                & \textbf{5}                                                & \textbf{6}   & \textbf{7}                                                & \textbf{8}                                                & \textbf{9}   & \textbf{10}                           & \textbf{\(\ldots\)} & \cellcolor[HTML]{FFCE93}\textbf{9 998}     & \cellcolor[HTML]{FFCE93}\textbf{9 999}   & \cellcolor[HTML]{FFCE93}\textbf{10 000}   \\
\textbf{\begin{tabular}[c]{@{}l@{}}Study series\\ size (\(t\))\end{tabular}} &                                               &                                                           &                                      &                                      &                                                           &                                                           &              &                                                           &                                                           &              &                                       &                     & \cellcolor[HTML]{FFCE93}                   & \cellcolor[HTML]{FFCE93}                 & \cellcolor[HTML]{FFCE93}                  \\ \cline{3-3} \cline{7-7} \cline{10-10}
\cellcolor[HTML]{A8ADF7}\textbf{1}                                           & \multicolumn{1}{l|}{\cellcolor[HTML]{A8ADF7}} & \multicolumn{1}{l|}{\cellcolor[HTML]{A8ADF7}\(z_{1, 1}\)} & \cellcolor[HTML]{A8ADF7}\(z_{1, 2}\) & \cellcolor[HTML]{A8ADF7}\(z_{1, 3}\) & \multicolumn{1}{l|}{\cellcolor[HTML]{C0C0C0}\(z_{1, 4}\)} & \multicolumn{1}{l|}{\cellcolor[HTML]{A8ADF7}\(z_{1, 5}\)} & \(z_{1, 6}\) & \multicolumn{1}{l|}{\cellcolor[HTML]{A8ADF7}\(z_{1, 7}\)} & \multicolumn{1}{l|}{\cellcolor[HTML]{C0C0C0}\(z_{1, 8}\)} & \(z_{1, 9}\) & \cellcolor[HTML]{A8ADF7}\(z_{1, 10}\) & \(\ldots\)          & \cellcolor[HTML]{B3B7F6}\(z_{1, 9 998}\)   & \cellcolor[HTML]{FFCE93}\(z_{1, 9 999}\) & \cellcolor[HTML]{B3B7F6}\(z_{1, 10 000}\) \\
\cellcolor[HTML]{A8ADF7}\textbf{2}                                           & \multicolumn{1}{l|}{\cellcolor[HTML]{A8ADF7}} & \multicolumn{1}{l|}{\cellcolor[HTML]{A8ADF7}\(z_{2, 1}\)} & \cellcolor[HTML]{A8ADF7}\(z_{2, 2}\) & \cellcolor[HTML]{A8ADF7}\(z_{2, 3}\) & \multicolumn{1}{l|}{\cellcolor[HTML]{C0C0C0}\(z_{2, 4}\)} & \multicolumn{1}{l|}{\cellcolor[HTML]{A8ADF7}\(z_{2, 5}\)} &              & \multicolumn{1}{l|}{\cellcolor[HTML]{A8ADF7}\(z_{2, 7}\)} & \multicolumn{1}{l|}{\cellcolor[HTML]{C0C0C0}\(z_{2, 8}\)} &              & \cellcolor[HTML]{A8ADF7}\(z_{2, 10}\) &                     & \cellcolor[HTML]{B3B7F6}\(z_{2, 9 998}\)   & \cellcolor[HTML]{FFCE93}                 & \cellcolor[HTML]{B3B7F6}\(z_{2, 10 000}\) \\  \cline{10-10}
\cellcolor[HTML]{CBCEFB}\textbf{3}                                           & \multicolumn{1}{l|}{\cellcolor[HTML]{CBCEFB}} & \multicolumn{1}{l|}{\cellcolor[HTML]{CBCEFB}\(z_{3, 1}\)} & \cellcolor[HTML]{CBCEFB}\(z_{3, 2}\) & \cellcolor[HTML]{CBCEFB}\(z_{3, 3}\) & \multicolumn{1}{l|}{}                                     & \multicolumn{1}{l|}{\cellcolor[HTML]{CBCEFB}\(z_{3, 5}\)} &              & \cellcolor[HTML]{CBCEFB}\(z_{3, 7}\)                      &                                                           &              & \cellcolor[HTML]{CBCEFB}\(z_{3, 10}\) &                     & \cellcolor[HTML]{CBCEFB}\(z_{3, 9 998}\)   & \cellcolor[HTML]{FFCE93}                 & \cellcolor[HTML]{CBCEFB}\(z_{3, 10 000}\) \\ \cline{3-3} 
\textbf{4}                                                                   &                                               &                                                           & \(z_{4, 2}\)                         & \(z_{4, 3}\)                         & \multicolumn{1}{l|}{}                                     & \multicolumn{1}{l|}{\(z_{4, 5}\)}                         &              & \(z_{4, 7}\)                                              &                                                           &              &                                       &                     & \cellcolor[HTML]{FFCE93}\(z_{4, 9 998}\)   & \cellcolor[HTML]{FFCE93}                 & \cellcolor[HTML]{FFCE93}\(z_{4, 10 000}\) \\
\textbf{5}                                                                   &                                               &                                                           & \(z_{5, 2}\)                         &                                      & \multicolumn{1}{l|}{}                                     & \multicolumn{1}{l|}{\(z_{5, 5}\)}                         &              &                                                           &                                                           &              &                                       &                     & \cellcolor[HTML]{FFCE93}\(z_{5, 9 998}\)   & \cellcolor[HTML]{FFCE93}                 & \cellcolor[HTML]{FFCE93}                  \\ \cline{7-7}
\textbf{6}                                                                   &                                               &                                                           & \(z_{6, 2}\)                         &                                      &                                                           & \(z_{6, 5}\)                                              &              &                                                           &                                                           &              &                                       &                     & \cellcolor[HTML]{FFCE93}\(z_{6, 9 998}\)   & \cellcolor[HTML]{FFCE93}                 & \cellcolor[HTML]{FFCE93}                  \\
\textbf{\(\ldots\)}                                                          &                                               &                                                           &                                      &                                      &                                                           &                                                           &              &                                                           &                                                           &              &                                       &                     & \cellcolor[HTML]{FFCE93}\(\ldots\)         & \cellcolor[HTML]{FFCE93}                 & \cellcolor[HTML]{FFCE93}                  \\
\textbf{136}                                                                 &                                               &                                                           &                                      &                                      &                                                           &                                                           &              &                                                           &                                                           &              &                                       &                     & \cellcolor[HTML]{FFCE93}\(z_{136, 9 998}\) & \cellcolor[HTML]{FFCE93}                 & \cellcolor[HTML]{FFCE93}                 
\end{tabular}
\end{table*}

\subsection{Accumulation Bias caused by Evidence-Based Research} \label{sec:EBR}
\begin{quote}
New research should not be done unless, at the time it is initiated, the questions it proposes to address cannot be answered satisfactorily with existing evidence. --Chalmers \& Glasziou (\citeyear{chalmers2009avoidable})
\end{quote}
In 2009, the term \lingform{Research Waste} was coined and this key recommendation was made. The recommendation further specifies that existing evidence should be obtained by a systematic review and summarized with a meta-analysis. But how exactly to answer the question whether new research is necessary or wasteful remained unclear. Nevertheless, the recommendation was important enough to be repeated, as was first done in an entire series on Research Waste with a specific recommendation on setting research priorities \cite{chalmers2014increase} and later in a paper that gave the recommendation its official name: \lingform{Evidence-Based Research} \cite{lund2016towards}. Support for these recommendations was provided by various retrospective cummulative meta-analyses that show how many studies were still performed while satisfactory evidence was already available. These cummulative meta-analysis judge \enquote{satisfactory evidence} based on a significance threshold, usually uncorrected for multiple testing (e.g. \citet{fergusson2005randomized}), which reminds us of the Accumulation Bias that occurs in living systematic reviews (Section \ref{sec:LSR}).

The larger consequence, however, is that Accumulation Bias is caused by any dependencies between results and series size and meta-analysis timing, and that Evidence-Based Research introduces such dependencies. Inspecting previous results to decide whether new research is necessary or wasteful therefore always introduces Accumulation Bias, whether it based on uncorrected or corrected thresholds. Also more subtle decision methods --- implicit rather than based on thresholds --- introduce Accumulation Bias, as was shown by \citet{kulinskaya2016sequential}. In fact, they describe the rationale behind their models --- among which the \lingform{power-law} model (Section \ref{sec:Kulinskaya}) --- as an example of bias introduced by guidelines to decide on \enquote{the usefulness of a new study} \enquote{with direct reference to existing meta-analysis.} \cite[p. 297]{kulinskaya2016sequential}.

So Evidence-Based Research causes bias, and our Accumulation Bias Framework demonstrates how it might affect the sampling distribution, whether based on explicit thresholds or implicit decision making. Does this mean that we cannot make Evidence-Based Research decisions to avoid research waste, while also controlling type-I errors? Fortunately, we do not need to be that pessimistic and can still embrace \lingform{Evidence-Based Research}. In Section~\ref{sec:LR} we show that tests based on likelihood ratios withstand Accumulation Bias and are very well suited to reduce research waste. But to do so, we first need to specify exactly what role is played by \lingform{time} in error control.

\section{\lingform{Time} in error control} 
Over time new study series are initiated, studies are added to existing study series and more meta-analyses are performed. To visualize how this process relates to error control, we need to start with a specific state of this expanding system. In 2001 an estimated minimum of 10 000 medical topics were covered in over half a million studies, thus requiring 10 000 meta-analyses if all were synthesized in a database such as the \lingform{Cochrane Database of Systematic Reviews}
 \cite{mallett2003many}. The number of studies in a series varied between 2 and 136, which we can use to describe the 2001 state of a possible database, that to be complete, also includes many unreplicated pilot studies. We could visualize this database in a table, with studies in the rows, topics in the columns and many missing entries. A sketch is shown in Table \ref{tab:Vis}.
 
 The conventional approach to error control, which we used to show the influence of \lingform{Gold Rush} Accumulation Bias in meta-analysis testing in Section \ref{sec:GoldTesting}, is a conditional approach. Since conventional meta-analysis does not raise any multiple testing issues, there is a hidden assumption that the timing of a meta-analysis \(\cA^{(t)}\) is independent from the data and each study series experiences only one meta-analysis. In Section \ref{sec:GoldAnaTimes} we took the \(t\) at which the sole meta-analysis is conducted to be either random or prespecified. This is shown in Table \ref{tab:Vis} by the black box enclosing the available studies on Topic 1. Other possible study series up for meta-analysis are shown by the boxes enclosing studies on Topic 5 and 8. Note that by assuming only one meta-analysis, a study series might continue growing but not be fully analyzed, as shown for Topic 5. 

In the conditional approach to error control, a three-study series \((Z_1, Z_2, Z_3)\) produces a possible draw from the \(Z^{(3)}\) sampling distribution. If we test our draw, the type-I error rate is defined as the fraction of \(t\)-study series that is considered significant if all \(t\)-study series were to be sampled from the null distribution. The question is: What study series are taken into account to specify this fraction? This is visualized in Table \ref{tab:Vis} by the dark blue and grey shading for \(t = 2\) and the dark blue and lighter blue shading for \(t = 3\). The unshaded topics and change of color between \(t = 2\) and \(t = 3\) show the flaw of this approach: some series might not survive up until a specific time \(t\), as for instance shown by the grey studies that are part of \(t = 2\) but not part of the error control for \(t = 3\). We also do not want every series to survive up until any arbitrary time \(t\) to avoid research waste \citep{chalmers2009avoidable}. The crucial point is that the series that do survive are no random sample from all possible \(t\)-study series. This is another illustration of Accumulation Bias such as the \lingform{Toy Story} scenario. The series deviates even more from the assumption of a random \(t\)-study draw if the meta-analysis time \(t\) is not random or prespecified, but dependent on the results, as expressed in Section \ref{sec:Timing}. We discuss the conventional conditional approach to meta-analysis error control in more detail in Section \ref{sec:Cond}.

The other possible approach to error control is surviving over analysis times, which means that it should be valid for any upcoming analysis time \(t\) within a series. So the probability that a type-I error --- ever --- occurs in the accumulating series is controlled, whether the series reaches a large size or not. This is visualized in Table \ref{tab:Vis} by the orange shading, and has a long run error rate that runs over series of any size, including the one-study series. This approach to error control is taken by methods for living systematic reviews such as \lingform{Trial sequential analysis} and \lingform{Sequential meta-analysis}. We discuss this approach of error control surviving over time in more detail in Section \ref{sec:Surv}.

\subsection{Error control conditioned on time}\label{sec:Cond}

The null distributions of the common/fixed meta-analysis \(Z\)-statistic shown in Figure \ref{fig:distrZ} are conditioned on the size of the series, which is the \lingform{time}: \(T \geq t\). We can use our Accumulation Bias framework to give this distribution a general description, where we use \(f_0(z^{(t)})\) to denote the assumed standard normal null distribution for the meta-analysis \(Z\)-score and obtain a conditional density using Bayes' rule:
\begin{equation} \label{phiCond}
\begin{split}
 f_0\left(z^{(t)}\,\middle|\, \cA^{(t)}, T \geq t \right) &= \frac{f_0(z^{(t)}) \cdot \ProbNull\left[\cA^{(t)}, T \geq t \,\middle|\, z^{(t)}\right]}{\ProbNull\left[\cA^{(t)}, T \geq t\right]} \\
 &= \frac{f_0(z^{(t)}) \cdot \overline{A}_0\left(t \,\middle|\, z^{(t)}\right)}{\overline{A}_0\left(t\right)}, \\
\text{where we define:} & \\
\overline{A}_0\left(t \,\middle|\, z^{(t)}\right) &:= \ExpNull\left[A\left(t \,\middle|\, Z_1, \ldots, Z_t\right) \,\middle|\, Z^{(t)} = z^{(t)}\right] \\
\overline{A}_0\left(t\right) &:= \ExpNull\left[A\left(t \,\middle|\, Z_1, \ldots, Z_t\right)\right], \\
\text{with under the equal} & \text{ study size assumption in (Eq. \eqref{Zt:Equaln})}\\
Z^{(t)} &= \frac{1}{\sqrt{t}} \sum_{i = 1}^{t} Z_{i}
\end{split}
\end{equation}
(extension to the general cases with unequal sample sizes is straightforward). For the \lingform{Gold Rush} example, \(\overline{A}_0\left(t\right)\) was given by Eq. \eqref{A(t)Gold Rush} and can be calculated if \(\omega\)s are known. \(\overline{A}_0\left(t\right)\) denotes the general probability of arriving at \(T \geq t\) under the null hypothesis, and so does  \(\overline{A}_0\left(t \,\middle|\, z^{(t)}\right)\), but with the restriction that we only take samples into account that result in meta-analysis score \(z^{(t)}\). The type-I error rates for the \lingform{Gold Rush} example shown in Table~\ref{tab:TypeIerrorTildeSim} are based on a randomly chosen or prespecified \(t\) for which \(\Prob[\cA^{(t)}] = 1\), and represent the following (with \(f_0\) as above in Eq. \eqref{phiCond}):

\begin{equation} \label{TypeICond}
\begin{split}
\ProbNull\left[\cE^{(t)}_{\textsc{type-I}} \,\middle|\, \cA^{(t)}, T \geq t \right] &= \int_{\minus \infty}^{\minus z_{\frac{\alpha}{2}}} f_0\left(z^{(t)} \,\middle|\, \cA^{(t)}, T \geq t \right) dz^{(t)} \\
& \quad + \int_{z_{\frac{\alpha}{2}}}^\infty f_0\left(z^{(t)} \,\middle|\, \cA^{(t)}, T \geq t \right) dz^{(t)}.
\end{split}
\end{equation}

\subsection{Error control surviving over time} \label{sec:Surv}

In living systematic reviews, a meta-analysis is performed after each new study (\(\Prob\left[\cA^{(t)}\right] = 1\) for all \(t\)). The properties on error control obtained by for example \lingform{Trial Sequential Analysis} are therefore surviving over analysis times \(t\) and depend on the joint distribution on the data and the maximum study series size \(T\). For \(\Prob\left[\cA^{(t)}\right]\) always 1, \(A(t) = S(t-1)\) and this joint distribution can be presented as follows:

\begin{equation}\label{eq:distrSurv}
\begin{split}
&f_0\left(z^{(1)}, \ldots, z^{(t)}, T = t \right)\\
& \quad = f_0\left(z^{(1)}, \ldots, z^{(t)}\right) \cdot \ProbNull\left[T = t \,\middle|\, z^{(1)}, \ldots, z^{(t)}\right],
\end{split}
\end{equation}
where we define
\begin{equation}
\begin{split}
\nonumber
& \quad \ProbNull\left[T = t \,\middle|\, z^{(1)}, \ldots, z^{(t)}\right] \\
& \quad \quad := \ExpNull\left[S(t-1 \,\middle|\, Z_1, \ldots, Z_{t-1}) \,\middle|\, Z^{(1)} = z^{(1)}, \ldots \right] \\
& \quad \quad \qquad - \ExpNull\left[S(t \,\middle|\, Z_1, \ldots, Z_t) \,\middle|\, Z^{(1)} = z^{(1)}, \ldots \right], \\
& \text{with under the equal study size assumption in (Eq. \eqref{Zt:Equaln})},\\
& \quad Z^{(t)} = \frac{1}{\sqrt{t}} \sum_{i = 1}^{t} Z_{i}, \\
& \text{ and with } f_0(z^{(0)}) = 1 \text{ and } \ProbNull\left[T \geq 1 \,\middle|\, z^{(0)}, z^{(1)}\right] = 1.
\end{split}
\end{equation}
The result \(\Prob[T = t] = S(t-1)-S(t)\) is known from survival analysis and made explicit in the Appendix \ref{sec:AppMaxTime}. When \(S(t)\) is known for all \(t\), it is possible to obtain error control that survives over analysis times \(T = t\) with thresholds \(z_{\frac{\alpha}{2}}^{(t)}\) that are functions of \(\alpha\), \(t\) and some \(T_{\text{max}}\) based on a maximum sample or information size. Such methods are known as \lingform{Trial sequential analysis} \citep{brok2008apparently, thorlund2008can, wetterslev2008trial} and \lingform{Sequential meta-analysis} \citep[Ch. 12]{whitehead2002meta} \citep{whitehead1997prospectively, higgins2011sequential}. If we assume a one-sided test, the approach to error control taken by these methods can be expressed as follows:

\begin{equation} \label{TypeIUncond}
\begin{split}
&\Exp_T\left[\ProbNull\left[\cE_{\textsc{type-I}}^{(T)} \,\middle|\, T\right]\right] \\
&\quad = \sum_{t = 1}^{T_{\text{max}}} \int_{z_{\frac{\alpha}{2}}^{(1)}}^\infty \ldots \int_{z_{\frac{\alpha}{2}}^{(t)}}^\infty f_0\left(z^{(1)}, \ldots, z^{(t)}, T = t \right) dz^{(1)} \ldots dz^{(t)} \\
&\quad = \alpha, \\ \\
& \text{with } f_0 \text{\ as above \eqref{eq:distrSurv}} \\
& \text{and } T = t \text{ only in the case } \lambda(t) = \mathbbm{1}_{Z^{(t)} \geq z^{(t)}_{\frac{\alpha}{2}}} = 1.
\end{split}
\end{equation}
The change in notation from \(T \geq t\) to \(T = t\) already hints at the limitations of this approach: the series size needs to be completely determined by the thresholds specified in the hazard function and nothing else. We discuss this limitation in more detail in the next section.

\subsection{Unknown and unreliable analysis time probabilities} \label{sec:Unknown}

To obtain thresholds to test \(z^{(t)}\) under Accumulation Bias, we need to know the probability \(A(t)\) (or only \(S(t)\)) for meta-analysis time \(t\). However, any of the scenarios described in Sections \ref{sec:Size} and \ref{sec:Timing} can be involved, and some can be influencing \(z^{(t)}\) simultaneously. Also, ethical imperatives might balance the bias, as illustrated by the following quote:

\begin{quote} 
A negative result will dampen enthusiasm and turn the attention of investigators to other possible protocols. A positive result will excite interest but may provide an ethical veto on further randomization. --Armitage (\citeyear{armitage1984controversies}) as cited by \citet{ellis2009temporal}
\end{quote}
We do not believe that the corrected thresholds \(z^{(t)}_{\frac{\alpha}{2}}\) from sequential methods like \lingform{Trial Sequential Analysis} can account for all Accumulation Bias, since they require very strict conformation to the stopping rule based on synthesized studies \(z^{(t)}\) and some have already argued that meta-analysts do not have such control over new studies \citep{chalmers1993meta}. \lingform{Sequential meta-analysis} was proposed for prospective meta-analyses \citep{whitehead1997prospectively, higgins2011sequential} and never intended for settings with retrospective dependencies. Stopping rules based solely on meta-analysis ignore dependencies that might already have arisen at the individual study level (such as in the \lingform{Gold Rush} example) and that meta-analyses might in practice not be performed continuously (so \(\Prob[\cA^{(t)}] \neq 1\) for some \(t\)). When meta-analyses are not performed continuously, as discussed in Section \ref{sec:Timing}, the specification of which series are included in the long run error control is missing (imagine for example that some of the columns 1, 2, 3 and 5 of meta-analyses in Table \ref{tab:Vis} be excluded in the long run error control because the individual study results were such that nobody will ever bother to perform a meta-analysis).

It might be very inefficient to try to avoid Accumulation Bias. As stated in the introduction, avoiding it would mean that results from earlier studies should be unknown when planning new studies as well as when planning meta-analyses (that is, the decision to do a meta-analysis after $t$ studies should not depend on the outcome of these studies). Achieving this might be impossible, since research is very often somehow inspired by other findings. Also, such approach 
cannot be reconciled with the \lingform{Evidence-Based Research} initiative to reduce waste.\citep{lund2016towards, chalmers2009avoidable,chalmers2014increase}.

We conclude that the Accumulation Bias process specifying \(A(t)\) can never be fully known and that avoiding an Accumulation Bias process will introduce more research waste. So we need a testing method that is valid regardless of the exact Accumulation Bias process. We will introduce such a method in Section \ref{sec:LR}, but first exhibit some evidence that, even though the recommendations from \lingform{Evidence-Based Research} still need renewed attention, Accumulation bias might already be at play.

\section{Intermezzo: evidence for the existence of Accumulation Bias} \label{sec:Evid}

\subsection{Agreement with empirical findings} \label{sec:EmpEvid}

Accumulation Bias arises due to dependencies in how a study series comes about (Section \ref{sec:Size}), and in the timing of the meta-analysis (Section \ref{sec:Timing}). We first discuss some indications of the former and then illustrate how these can be reinforced by some approaches to the latter.

If citations of previous results are a real indication of why a replication study is performed, than many such dependencies have been demonstrated in the literature on \lingform{reference/citation bias} \citep{gotzsche1987reference, egger1998bias}. Citation or reference bias indicates that initial satisfactory results are more often cited than unsatisfactory results, thus some sort of \lingform{Gold Rush} occurs. Studies into citations indicate that early small trials are much more often cited than later large trials (e.g. \citet{fergusson2005randomized, robinson2011systematic}), which might limit the \lingform{Gold Rush} to the early studies in a series, such as indicated by \citet{whitehead2002meta} (Section \ref{sec:Whitehead}). Many studies have found that early studies are unreliable predictors of later replications in a study series \citep{roberts2015systematic, chalmers2016systematic} (and see references 6-34 in \citet{ioannidis2008most} and references 33-49 in \citet{pereira2011statistically}), which is also an indication of early study Accumulation Bias.

Other empirical findings suggest that Accumulation Bias might occur throughout a series, but to a lesser extend in later studies. \citet{gehr2006fading}, for example, report effect sizes that decrease over time, but in which study size did not play a significant role. What has been recognized as \lingform{regression to the truth} in heart failure studies, might also be characterized as Accumulation Bias \citep{krum2003phase}. But this effects will be difficult to limit to only a few early studies, so excluding a certain number from meta-analysis, as proposed in \citet[p. 197]{whitehead2002meta} (Section \ref{sec:Whitehead}), might therefore be a too crude measure. 

The Proteus effect \citep{pfeiffer2011quantifying, ioannidis2005early, ioannidis2005contradicted} describes how early replications can be biased against initial findings. If early contradicting findings spur a large series of studies into a phenomenon, it introduces a more complex pattern of Accumulation Bias that does not have a straightforward dominating direction. The same holds for the \lingform{Value of Information} approach, to decide on replication studies \citep{claxton2006using, claxton2002rational}.

There is quite some literature with suggestions on when a meta-analysis should be updated. One general recommendation is to do so when studies can be added that will have a large effect on the meta-analysis \citep{moher2006systematic, moher2007systematic, moher2008and}. If such recommendations reflect an overall tendency in timing of meta-analysis, Accumulation Bias might be re-enforced by the timing of the meta-analysis: initial misleading studies might have spurred a study series, and might also indirectly encourage a meta-analysis after later studies report deviating results.

\subsection{Agreement with intuitions about priors}

The famous paper \enquote{Why Most Published Research Findings are False} \citep{ioannidis2005most} introduced the concept of field specific prior odds to a large audience. The prior odds were presented as the \enquote{Ratio of True to Not-True Relationships (\(R\))}, which has the same meaning as the fraction of pilot studies from the null and alternative distribution (\(\pi/(1 - \pi)\)) in the terminology of this paper. \citet{ioannidis2005most} combines this ratio with the average power and type-I error of tests in a research field to obtain a field-specific estimate of the Positive Predictive Value (\lingform{PPV}) of a significant result. This is the expected rate or target rate of true to false rejections, and the same as \(\gamma \cdot \pi/(1 - \pi)\) in Section \ref{sec:LRcondition} of this paper.

\citet{ioannidis2005most} provides prior odds of various research fields and publication types for which two are of interest to Accumulation Bias: \enquote{Adequately powered RCT with little bias} and \enquote{Confirmatory meta-analysis of good-quality RCTs}. For the first of these an \(R\) of 1:1 is provided and for the second an \(R\) of 2:1. So a distinction is made between topics worthy of only one individual study and those that evoke a series of studies eligible for meta-analysis.

How would the researchers involved in replicating RCTs know that their topic is worthy of a series of studies in comparison to just one? The difference between prior odds of the two indicates that this is no random decision. The only available source of information would be previous study results, hence introducing dependence between study series size and study results: Accumulation Bias. So the prior odds \(R\) specified by \citet{ioannidis2005most} is actually \(\frac{\pi \cdot \overline{A}_1(t)}{(1 - \pi) \cdot \overline{A}_0(t)}\), with \(\overline{A}_1(1) = 1\) and \(\overline{A}_0(1) = 1\) for primary studies.

\section{Likelihood ratios' independence from meta-analysis time} \label{sec:LR}
In Section \ref{sec:Unknown} we argued that any approach to model the analysis time probabilities \(A(t)\) is unreliable: in realistic and practically relevant scenarios, the ingredients required to calculate $A(t)$ will be unknown. Therefore, we need to define test statistics that are independent from how a series size or meta-analysis comes about. A possible form of such a test statistic is the likelihood ratio, which we discuss from the two approaches to error control: in the next section \ref{sec:LRcondition} from the perspective of error control conditioned on time, and in Section \ref{sec:LRsurviving} from the perspective of error control surviving over time.

Our proposed use of the likelihood ratio is based on the following extraordinary 
property\footnote{This property is related to the well-known fact that the Bayesian posterior based on data, 
when the priors are determined independently of the sample size, takes on the same value irrespective of the stopping rule that gave rise to the observations 
\citep{hendriksen2018optional}}, already recognized by \citet{berger1988relevance} and shown in Eq. \eqref{LR}: The likelihood ratio is a test statistic that depends on the specification of some alternative distribution \(f_1\). Any data sampled from an alternative distribution will have the same analysis time probabilities as data sampled from the null distribution, since analysis time probabilities are independent from the data-generating hypothesis (Section \ref{sec:AnaTimeInd}). When a likelihood ratio statistic is obtained for known data, the analysis time probability is a constant factor that is the same in the numerator and denominator of the likelhood ratio and therefore drops out of the equation:
\begin{equation} \label{LR}
\begin{split}
& \LR^{(t)}\left(z_1, \ldots, z_t , \cA^{(t)}, T \geq t \right) \\
& \quad := \frac{f_{1}\left(z_1, \ldots, z_t \right) \cdot \ProbAlt( \cA^{(t)}, T \geq t \mid z_1, \ldots, z_t )}{f_{0}\left(z_1, \ldots, z_t \right) \cdot \ProbNull( \cA^{(t)}, T \geq t \mid z_1, \ldots, z_t ) } \\
& \quad = \frac{f_{1}\left(z_1, \ldots, z_t \right) \cdot A\left(t \,\middle|\, z_1, \ldots, z_t\right)}{f_{0}\left(z_1, \ldots, z_t \right) \cdot A\left(t \,\middle|\, z_1, \ldots, z_t \right)} \\
& \quad = \frac{f_{1}\left(z_1, \ldots, z_t \right)}{f_{0}\left(z_1, \ldots, z_t \right)} \\
& \quad = \LR\left(z_1, \ldots, z_t \right).
\end{split}
\end{equation}
Here we used the standard definition of likelihood ratio for the case that the likelihood jointly involves continuous-valued data and discrete events, and we critically used the fact that the probability of $ \cA^{(t)}, T \geq t$ does not depend on whether the null or the alternative distribution generated the data.

In the following two sections we discuss two means of using likelihood-ratio based tests that yield results that are valid irrespective of accumulation bias.\footnote{To avoid any confusion, let us highlight that our likelihood-ratio based tests are {\em never\/} equivalent to $p$-value based tests. While some $p$-value based tests (such as the Neyman-Pearson most powerful test) can be written as likelihood ratio tests, these are invariably of the form `reject at significance level $\alpha$ if $\LR(z_1,\ldots, z_t) \geq \gamma$ where $\gamma$ is chosen such that $\ProbNull(f_1(z_1, \ldots, z_t)/f_0(z_1,\ldots, z_t) \geq \gamma) = \alpha$. In contrast, we choose $\gamma$ in a way that does not depend on knowledge of the tail area under $\ProbNull$ (e.g. in Section~\ref{sec:LRsurviving} we take $\gamma =1/\alpha$, and there the equality above is a (strict) inequality).}

\subsection{Likelihood ratio's error control conditioned on time} \label{sec:LRcondition}
A large study series has an extremely low probability of occurring under the null hypothesis in the \lingform{Gold Rush} scenario, and under any other similar Accumulation Bias setting. The probability of reaching a certain study series size \(t\) is much larger under any alternative hypothesis when the power of the test for that alternative hypothesis (\(1-\beta\)) is larger than the type-I error \(\alpha\). Due to this fact, it is possible to control an error rate if we assume that a certain fraction of pilot studies (or topics, see Table \ref{tab:Vis}) \(\pi\) are sampled from the alternative distribution and a proportion \((1-\pi)\) of pilot studies from the null. This way, we are able to control the fraction of true rejections \(1 - \ProbAlt\left[\cE^{(t)}_{\textsc{type-II}} \,\middle|\, \cA^{(t)}, T \geq t \right]\) (complement of type-II errors) to false rejections \(\ProbNull\left[\cE^{(t)}_{\textsc{type-I}} \,\middle|\, \cA^{(t)}, T \geq t \right]\). 

We can achieve such error control conditioned on time --- e.g. error control taking into account only \(t\)-study meta-analyses --- if we define thresholds based on the \lingform{Bayes posterior odds}, which, by Bayes' theorem, are given by $O_{\text{post}}\left(z_1, \ldots, z_t \right) = \LR\left(z_1, \ldots, z_t \right) \cdot \frac{\pi}{1 - \pi}$. Remarkably, these are not affected by the mechanism underlying the decisions to continue studies or perform meta-analyses:

\begin{equation} \label{postOdds}
\begin{split}
& O_{\text{post}}\left(z_1, \ldots, z_t \,\middle|\, \cA^{(t)}, T \geq t\right) \\
& \quad := \frac{\Prob\left[H_1 \,\middle|\, z_1, \ldots, z_t, \cA^{(t)}, T \geq t\right]}{\Prob\left[H_0 \,\middle|\, z_1, \ldots, z_t, \cA^{(t)}, T \geq t\right]} \\
& \quad = \frac{f_1\left(z_1, \ldots, z_t , \cA^{(t)}, T \geq t\right) \cdot \pi}{f_0\left(z_1, \ldots, z_t , \cA^{(t)}, T \geq t\right) \cdot (1 - \pi)}\\
& \quad = \LR^{(t)}\left(z_1, \ldots, z_t , \cA^{(t)}, T \geq t\right) \cdot \frac{\pi}{1 - \pi}\\
& \quad = \LR\left(z_1, \ldots, z_t \right) \cdot \frac{\pi}{1 - \pi} \\
& \quad = O_{\text{post}}\left(z_1, \ldots, z_t \right).
\end{split}
\end{equation}
We can set a threshold \(\gamma\) based on the rate of true to false rejections, so \(\gamma = 16\) would mean that we try to achieve 16 times as many true rejections than false rejections \(\gamma = \frac{1-\beta}{\alpha}\), which is the the usual goal of a primary analysis with intended power \(1-\beta = 0.8\) and type-I error rate \(\alpha = 0.05\). To obtain error control, we need to specify the \lingform{pre-experimental rejection odds} \citep{bayarri2016rejection} \(\gamma \cdot \frac{\pi}{1-\pi}\) and use these to threshold the posterior odds (Eq. \eqref{postOdds}). We define \(R\) to be the region of the sample space and \(\cR\) the event for which \(O_{\text{post}}(z_1, \ldots, z_t) \geq \gamma \cdot \frac{\pi}{1-\pi}\), i.e. the event that we reject, and obtain the following:

\begin{equation} \label{postOddsThreshold}
\begin{split}
&\frac{1 - \ProbAlt\left[\cE^{(t)}_{\textsc{type-II}} \,\middle|\, \cA^{(t)}, T \geq t \right]}{\ProbNull\left[\cE^{(t)}_{\textsc{type-I}} \,\middle|\, \cA^{(t)}, T \geq t \right]} \\
& \quad = \frac{\ProbAlt\left[O_{\text{post}}\left(Z_1, \ldots, Z_t \,\middle|\, \cA^{(t)}, T \geq t\right) \geq \gamma \cdot \frac{\pi}{1-\pi}\right]}{\ProbNull\left[O_{\text{post}}\left(Z_1, \ldots, Z_t \,\middle|\, \cA^{(t)}, T \geq t\right) \geq \gamma \cdot \frac{\pi}{1-\pi} \right]} \\
& \quad = \frac{\ProbAlt\left[O_{\text{post}}(Z_1, \ldots, Z_t) \geq \gamma \cdot \frac{\pi}{1-\pi}\right]}{\ProbNull\left[O_{\text{post}}(Z_1, \ldots, Z_t) \geq \gamma \cdot \frac{\pi}{1-\pi} \right]} \\
& \quad = \frac{\ProbAlt[\cR]}{\ProbNull[\cR]} \geq \frac{\ProbAlt[\cR]}{\ProbAlt[\cR] \cdot \frac{1}{\gamma}} = \gamma,
\end{split}
\end{equation}
where the inequality follows since if \\
\(O_{\text{post}}\left(z_1, \ldots, z_t \,\middle|\, \cA^{(t)}, T \geq t\right) \geq \gamma \cdot \frac{\pi}{1-\pi}\):
\begin{equation}
\begin{split}
\frac{f_1\left(z_1, \ldots, z_t\right)}{f_0\left(z_1, \ldots, z_t\right)} \cdot \frac{\pi}{1 - \pi} &\geq \gamma \cdot \frac{\pi}{1-\pi} \\ 
\text{then} \quad \frac{f_1\left(z_1, \ldots, z_t\right)}{f_0\left(z_1, \ldots, z_t\right)} &\geq \gamma \quad \text{and} \\
 \ProbNull[\cR] = \int_{R} f_0(z_1, \ldots, z_2) &\leq \int_{R} \frac{f_1(z_1, \ldots, z_2)} {\gamma} = \frac{\ProbAlt[\cR]}{\gamma}.
\end{split}
\end{equation}
So by specifying \(\frac{\pi}{1-\pi}\) and an intended rate of true to false rejections \(\gamma\), we can calculate the posterior odds based on the likelihood ratio, compare it to the threshold based on \(\gamma\) and control fraction \(\gamma\) of type-I errors under the null hypothesis. Note that any \(\cA^{(t)}\) is allowed, also multiple testing in a series or selection for the most promising meta-analysis timing. Setting a threshold to the Bayes posterior odds as described above, achieves conditional error control under any form of Accumulation Bias.

\subsection{Likelihood ratio's error control surviving over time} \label{sec:LRsurviving}
A likelihood ratio itself can be used as a test statistic to obtain a procedure that controls \(\ProbNull[\cE_{\textsc{type-I}}]\) surviving over analysis times \(t\), as in Section \ref{sec:Surv}. Suppose we simply reject if the likelihood ratio in favor of the alternative is larger than $1/\alpha$, ignoring any knowledge we might have about the accumulation bias process and the prior odds. We then find: 
\begin{equation}\label{eq:lifeline}
\begin{split}
& \ProbNull\left[
\text{there exists $t \leq T$ with\ }
\cE^{(t)}_{\textsc{type-I}} \text{\ and\ } \cA^{(t)} \right] \\
& \quad = \ProbNull\left[
\exists t \leq T:\;
\cE^{(t)}_{\textsc{type-I}} ; \cA^{(t)} \right] \\
& \quad = \ProbNull\left[
\exists t \leq T:\; 
\LR^{(t)}\left(Z_1, \ldots, Z_t \right) \geq \frac{1}{\alpha} ; 
 \cA^{(t)} 
\right] \\
& \qquad \leq \ProbNull\left[
\exists t > 0:\;
\LR^{(t)}\left(Z_1, \ldots, Z_t \right) \geq \frac{1}{\alpha} 
\right] \leq {\alpha}.
 \end{split}
\end{equation}
The final inequality is a classic result, proofs of which can be found in, for example, \citet{robbins1970statistical,shafer2011test} and (with substantial explanation) \citet{hendriksen2018optional}; see also \citet{royall2000probability}.

Thus, the type-I error control survives over time in the sense that the $\ProbNull$-probability that we {\em ever\/} reject at a meta-analysis time is bounded by $\alpha$. To further illustrate and interpret error control surviving over time, we define $$\cF^{(t)}_{\textsc{type-I}} = \cE^{(t)}_{\textsc{type-I}} \cap \overline{\cE}^{(t-1)}_{\textsc{type-I}},
\cap \ldots \cap \overline{\cE}^{(1)}_{\textsc{type-I}}$$ 
as the event that the {\em first\/} type-I error \(\cE^{(t)}_{\textsc{type-I}}\) in a series happens at time \(t\) (here $\overline{\cE}^{(t')}_{\textsc{type-I}}$ means `no type-I error at time $t'$). As we show in Appendix~\ref{sec:AppFromSumToExists}, the previous inequality implies that 
\begin{equation}\label{eq:deadline}
\sum_t \ProbNull\left[\cF^{(t)}_{\textsc{type-I}}, \cA^{(t)}, T \geq t \right] \leq \alpha.
\end{equation}
The change in notation from $\cE^{(t)}_{\textsc{type-I}}$ to $\cF^{(t)}_{\textsc{type-I}}$ is necessary since we want a general result for all forms of Accumulation Bias and do not want to assume that the series stops growing after the threshold is crossed (as is assumed in living systematic reviews, see Section \ref{sec:LSR}). But since it is not possible to control the amount of errors if multiple errors are made in the same series, we count only the first error in Eq. \eqref{eq:deadline}. As such, we are able to control the number of topics for which an error ever occurs in the series by comparing the likelihood ratio to the threshold \(\frac{1}{\alpha}\).

It may seem surprising that it is possible to obtain error control in the sense of Eq. \eqref{eq:deadline} for Accumulation Bias scenarios like \lingform{Gold Rush} example. After all, in this example large study series have only a large probability to occur if they contain many extreme (significant) results. So it seems that we would inevitably hit a type-I error once we perform a meta-analysis. But note that in this example, the expectation of \(A(t \mid Z_1, \ldots, Z_t)\) (\(\overline{A}_0(t)\)) is much larger for small \(t\) --- due to the \(S(t)\) component --- so that most meta-analyses will be of small study series, or even one-study series, with small type-I error rates. In terms of Table~\ref{tab:Vis}, controlling error this way is possible because error control runs over all topics, regardless of the realized series size. Thus, such error control is only meaningful if the series for each topic are continuously monitored --- including those consisting of only pilot studies.

\section{The choice between error control conditioned and surviving over time} \label{sec:Choice}

Many meta-analysts seem reluctant to apply living systematic review techniques to all meta-analyses. We believe that this reluctance can be defended based on the assumed approach to error control surviving over time. Surviving over time means that all possible analysis times are weighted and that --- in the long run ---- a large proportion of meta-analyses will be one-, two- and three-study meta-analyses
and never expand. To the occasional meta-analyst, not involved in continuously updating meta-analyses, two- or three-study meta-analyses might never occur. Also, it requires a stretch of mind to imagine one-study meta-analyses part of the long run properties of your specific 15-study meta-analysis. But it has been argued that \enquote{primary research is increasingly viewed as part of a wider sequential process} \citep[p. 918]{higgins2011sequential}, or at least, that it should be \cite{lund2016towards}. Whether this approach to error control is acceptable might also be very field specific. Among medical meta-analyses in the Cochrane Database of Systematic Reviews, two- and three-study meta-analyses are common \cite{davey2011characteristics}, but in other fields meta-analyses might only be performed if many more studies are available.

If, on the other hand, we want to stick to the conventional conditional approach to meta-analysis, we need additional assumptions on the fraction \(\pi\) of true alternative hypotheses among pilot studies to threshold the posterior odds. Assuming a base rate \(\pi\) means that we are essentially Bayesian about the null and alternative hypothesis\footnote{We do not necessarily have to be {\em completely\/} Bayesian: even if the null and/or alternative are composite, we can define \enquote{likelihood ratios} that do not rely on prior guesses about the parameters within the models. But we do need to be partially Bayesian, in the sense that we need to specify a base rate for the null \citep{GrunwaldHK19}}, but there is no need to be strictly Bayesian: in practice, we might play around, and try best case and worst case \(\pi\), to see how it affects our posterior odds. The important thing for us to note within the context of this paper is that, when concentrating on posterior odds, we can ignore all details of the Accumulation Bias process and still obtain meaningful results, in the form of error control that balances type-I and type-II errors.

Summarizing: If we prefer conditional error control, we can obtain meaningful error control despite Accumulation Bias if we use tests based on likelihood ratios, but using prior odds for the base rates (and being partially Bayesian) is then unavoidable. If we prefer not to rely on any prior odds, we can still obtain meaningful error control despite Accumulation Bias if we use tests based on likelihood ratios, but then we have to resort to error control surviving over time instead of conditional error control.

The former, conditional approach balances type-I and type-II errors and thus takes power into account. The importance of taking power (the complement of a the type-II error rate) into account has been argued before by many \citep{simmonds2017living}. In the general approach to error control in individual studies, the expected type-I error rate is fixed by the significance level \(\alpha\), and the type-II error rate minimized by the experimental design and sample size. In retrospective meta-analysis, however, sample size (or study series size \(t\)) is not under the control of the meta-analyst. Also, the study series size \(t\) is only a snapshot of a possibly growing series (\(T \geq t\)), since more studies might be performed in the future. Therefore also estimations of meta-analysis power are snapshots at a specific meta-analysis time. Nevertheless, it is often argued that many meta-analyses are underpowered \citep{turner2013impact, davey2011characteristics} and that this should be taken into account in evaluating significance in meta-analyses. In Trial Sequential Analysis \citep{wetterslev2008trial} for example, an alternative hypothesis is formulated to judge the fraction of a required sample size available at \(t\) studies. A later review on trial sequential analysis noted:

\begin{quote}
statistical confidence intervals and significance tests, relating exclusively to the null hypothesis, ignore the necessity of a sufficiently large number of observations to assess realistic or minimally important intervention effects. --Wetterslev, Jakobsen \& Gluud (\citeyear[p. 12]{wetterslev2017trial})
\end{quote}
Testing procedures based on likelihood ratios are very well suited to take an alternative distribution with minimally important intervention effect into account. Especially when balancing type-I error and power by thresholding posterior odds. Specifying power in tests without fixed sample sizes is studied extensively in \citet{GrunwaldHK19} and will be the focus of future research into likelihood ratios for meta-analysis.

\section{Why likelihood ratios work: dependencies as strategy} \label{sec:Betting}

We calculate p-values to judge the extremeness of our results under the null hypothesis, and to control type-I errors. But the p-value method is a fairly complicated approach to that goal when it comes to meta-analysis: To obtain a valid p-value for a series of studies, the sampling distribution under the null hypothesis needs to specify exactly how the series and the meta-analysis timing came about. Only for a completely and accurately specified process can the extremeness of the data be judged and compared to a threshold based on the tail area of the sampling distribution.

Fortunately, much simpler approaches to the same goal can be found. One intuitive way is to consider a series of bets \(s(Z_1), s(Z_2), \ldots, s(Z_t)\) against the null hypothesis that make a profit when observed study results are extreme. The more extreme the results, the larger the profit. The bet needs to be designed in such a way that, under the null hypothesis, no profit is to be expected. Each null result might costs \$1 to play the bet, but in expectation also makes a \$1 profit:

\begin{equation}\label{eq:sval}
\ExpNull[s(Z_t)] = \$1.
\end{equation} 
Suppose that you start by investing \$1 in the first bet. After each study, you either decide to do a new study, and reinvest all profit obtained so far, or to stop and cash out. If you cash out after, for example, three studies, your profit is $s(Z_1) \cdot s(Z_2) \cdot s(Z_3)$. 

As long as Eq. \eqref{eq:sval} holds for each bet, you cannot expect to profit under the null hypothesis; no matter what the process is for deciding, based on past data, to continue to new studies or to stop. This can be mathematically proven using martingale theory, but intuitively the reason is clear: The situation is entirely analogous to that in a casino where you cannot expect to make a salary out of playing --- no matter how sophisticated the strategy you use on the order of the games or when you want to play or want to go home. Thus, irrespective of the rules used for continuation and stopping, making a large profit casts doubt on the null hypothesis even without knowledge of the entire sampling distribution.

This idea of testing by betting is described in great detail by \citet{shafer2019game}, and \citet{shafer2011test} show that a likelihood ratio is a beautiful way to specify such bets. Briefly, if we set $s(Z_t) = f_1(Z_t)/f_0(Z_t)$, then Eq. \eqref{eq:sval} obviously holds:

\begin{equation}
\ExpNull\left[\frac{f_1(Z_t)}{f_0(Z_t)}\right] = \int_z f_0(z) \frac{f_1(z)}{f_0(z)}dz = \int_z f_1(z)dz = 1.
\end{equation}
Under this definition, $s(z_1) \cdot \ldots \cdot s(z_t)$ has two interpretations: First, it is the joint likelihood ratio for the first $t$ studies. Second, it is the amount of profit made by sequentially reinvesting in a bet that is not expected to make a profit under the null hypothesis. 

So we can think of the meta-analyst acting at time $t$ as earning the profit specified by the likelihood ratio of the data until the $t$-th study, and using that information to advise on reinvestment in future studies. This procedure will not lead to bankruptcy if the null hypothesis is true, and will therefore allow you to keep reinvesting. If the null hypothesis is not true, the better the focus of the bets --- determined by how close the alternative distribution in the likelihood ratio is to the data-generating distribution --- the larger the expected profit. The crucial point is that every strategy is allowed, so also the ineffective ones that produce research waste: also not taking into account earlier studies is a strategy.

This interpretation --- likelihood ratios as betting strategies --- explains how dependencies in the series relate to the test statistic. Any Accumulation Bias process can be considered a strategy to reinvest profit made so far, by deciding on new studies (\(S(t)\)), or cashing out the current profit (equivalent to performing a meta-analysis at time $t$ and advising against further studies: \(\cA^{(t)}, T = t\)). This is the intuition behind the proof of results like Eq. \eqref{eq:lifeline} and \eqref{eq:deadline} --- bounds on type-I error probability in meta-analysis ---- that can be derived without knowledge of the Accumulation Bias process. These bounds simply express that under the null, a large profit is unlikely under the null no matter what the Accumulation Bias is.

\begin{quote}
it is always legitimate to continue betting, and this makes each individual study a more informative element of a research program or a meta-analysis -- Shafer (\citeyear[p. 2]{shafer2019language})
\end{quote}

In contrast to an all-or-nothing test for one study, inspecting the betting profit of a study is a way to test the data without loosing the ability to build on it in future studies. The likelihood ratio has the ability to maximize the rate of growth among all studies in a series, instead of the power of a single p-value test on a prespecified series size or stopping rule \cite{shafer2019language}. It allows for promising but inconclusive initial studies and small study series to be revisited in the light of new studies, but also to keep track of the combined evidence at any time.

In this sense, the use of likelihood ratios in meta-analysis is a statistical implementation of the goals of the \lingform{Evidence Based Research Network} \citep{lund2016towards}. Choosing your bets wisely, by informing new studies by previous results is just another betting strategy. You optimize what studies to perform, and how to design and analyze them. Implementing this rationale in the statistics allows to maximize the efficiency of future research and reduce research waste \citep{chalmers2009avoidable}.

\subsection{Expanding likelihood ratios to \lingform{Safe Tests}}

When the null hypothesis is simple, it can be shown that either using bets that satisfy Eq. \eqref{eq:sval} under the null or using likelihood ratios or using Bayes factors is equivalent, and the gambling approach can be viewed as a form of Bayesian inference. But for composite null (as in the $t$-test scenario, with unknown variance $\sigma^2$), the situation is trickier: bets that satisfy Eq. \eqref{eq:sval} under all distributions in the null hypotheses can still be constructed, but their relation to likelihood ratios is more complicated. The paper {\em Safe Testing\/} \cite{GrunwaldHK19} investigates this setting in great detail and shows that `error control surviving over time' \ (Section~\ref{sec:LRsurviving}) can still be obtained for general composite null.

\section{Discussion} \label{sec:Discussion}

We need to consider \lingform{time} --- study chronology and analysis timing --- in meta-analysis. We need it because estimates are biased by Accumulation Bias when they assume that a \(t\)-study series is a random sample from all possible \(t\)-study series, while in fact dependencies arise in accumulating science. We also need \lingform{time} because sampling distributions are greatly affected by it, and the (p-value) tail area approach to testing is very sensitive to the shape of the sampling distribution. And we need to consider \lingform{time} because it allows for new approaches to error control that recognize the accumulating nature of scientific studies. Doing so also illustrates that available meta-analysis methods --- general meta-analysis and methods for living systematic reviews --- target two very different approaches to type-I error control.

We believe that the exact scientific process that determines meta-analysis time can never be fully known, and that approaches to error control need to be trustworthy regardless of it. A likelihood ratio approach to testing solves this problem and has even more appealing properties that we will study in a forthcoming paper. Firstly, it agrees with a form of the stopping rule principle \citep{berger1988relevance}. Secondly, it agrees with the \lingform{Prequential principle} \cite{dawid1984present}. Thirdly, it allows for a betting interpretation \cite{shafer2019game, shafer2019language}: reinvesting profits from one study into the next and cashing out at any time.

But this approach still leaves us with a choice: either assume a prior probability \(\pi\) and separate meta-analysis of various sizes from each other and individual studies, or control the type-I error rate over all analysis times \(t\) and include individual studies in the meta-analysis world. The first approach is more of a reflection of the current reality in meta-analysis, while the second can be aligned with the goals from the \lingform{Evidence-Based Research Network} \citep{lund2016towards} and \lingform{living systematic reviews} \citep{simmonds2017living}.

Accumulation Bias itself might not need to be corrected at all, which is why we want to close this paper with the following quote:

\begin{quote}the intuitive notion that bias is something bad which must be corrected for, does not even fit well within the frequentist framework. [...] one could not state \enquote{use estimate \(\overline{X}\) for a fixed sample size experiment, but use \(\overline{X} - c (\overline{X})\) (correcting for bias) for a sequential experiment,} and retain frequentist admissibility in the \enquote{real} situation where one encounters a variety of both types of problems. The requirement of unbiasedness simply seems to have no justification.
--Berger \& Berry (\citeyear[p. 67]{berger1988relevance})
\end{quote}

\subsection*{Data availability}

\subsubsection*{\textit{Underlying data}}

All data underlying the results are available as part of the article and no additional source data are required

\subsubsection*{\textit{Extended data}}

See Appendix \ref{sec:AppCode} for description of simulation and visualization R code and packages used to generate the code. Code is available from Electronic Archiving System - Data Archiving and Networked Services (EASY -DANS)\\

EASY-DANS: Accumulation Bias in Meta-Analysis: The Need to Consider Time in Error Control. \url{https://doi.org/10.17026/dans-x56-qfme} \cite{EASY-DANS} \\

Data are available under the terms of the \href{https://creativecommons.org/publicdomain/zero/1.0/}{Creative Commons Zero "No rights reserved" data waiver} (CC0 1.0 Public domain dedication).

\subsection*{Grant information}
This work is part of the NWO TOP-I research programme {\em Safe Bayesian Inference\/} [617.001.651], which is financed by the Netherlands Organisation for Scientific Research (NWO).

\subsection*{Acknowledgements}
This paper benefited from discussions with Allard Hendriksen, Rosanne Turner, Muriel Pérez, Alexander Ly and Glenn Shafer.


{\small\bibliographystyle{apalike}
\bibliography{References_Accumulation_Bias_paper.bib}}

\newpage
\onecolumn

\appendix
\input{Appendix}

\end{document}

%% file: Appendix.tex
\section{Appendix}

\subsection{Common/fixed-effect meta-analysis} \label{sec:AppCommFixMetaAnalysis}

Here we derive Eq. \eqref{Zt:Unequaln} and \eqref{Zt:Equaln}, shown in \eqref{Ztderiv}, from the notation in \cite{borenstein2009introduction}, specifically for the setting where means and standard deviations are reported in the study series \citet[Ch. 4 ]{borenstein2009introduction}. We slightly adjusted the notation by using \(\overline{X}_{\textsc{t}}\) and \(\overline{X}_{\textsc{p}}\) instead of \(\overline{X}_1\) and \(\overline{X}_2\) to indicate the treatment and placebo group estimate --- to avoid confusion with the study numbering --- and using \(\overline{D_{i}}\) instead of \(D_i\) \cite[p. 22]{borenstein2009introduction} or \(Y_i\) \cite[p. 66]{borenstein2009introduction} as an analogy to the group study mean \(X_i\) and we denote its standard deviation as \(\sigma_{D_{i}}\). We introduce the superscript \(^{(t)}\) to emphasize a meta-analysis estimate of a series of studies \(1\) up to \(t\).\\

Let \(D_{i} = X_{\textsc{t}i} - X_{\textsc{p}i}\) be a random variable that denotes the difference between two observations (random or paired) from the treatment group (\(X_{\textsc{t}i}\)) and the placebo group (\(X_{\textsc{p}i}\)) in study \(i\). Let \(\hat{\sigma}_{D_{i}}\) be the estimate of the population standard deviation of these difference scores in study \(i\). Following the usual assumptions of common/fixed-effect meta-analysis, no distinction is made between \(\hat{\sigma}_{D_{i}}\) and the true \(\sigma_{D_{i}}\) \cite[p. 264]{borenstein2009introduction} and for simplicity, we assume these standard deviations to be equal across studies:

\begin{equation} \label{knownSigma}
\text{For all } i, j \in \{1,2,\ldots,t\} \quad \hat{\sigma}_{D_{i}} = \sigma_{D_{i}} = \hat{\sigma}_{D_{j}} = \sigma_{D_{j}} = \sigma_{D}
\end{equation}
Let \( \overline{D}_{i} = \overline{X}_{\textsc{t}i} - \overline{X}_{\textsc{p}i} \) be the estimated treatment effect in study \(i\), i.e. the difference between the average effect in the treatment group \(\overline{X}_{\textsc{t}i}\) in study \(i\) and the average effect in the placebo group \(\overline{X}_{\textsc{p}i}\) in study \(i\). The population treatment effect is denoted by \(\Delta\), and is the difference between the population mean effects in the two groups, \(\Delta = \mu_{\textsc{t}} - \mu_{\textsc{p}}\) \cite[p. 21]{borenstein2009introduction}. Let \(Z_{i} = \frac{\overline{D}_{i}}{SE_{\overline{D}_{i}}}\) be the treatment \(Z\)-score of study \(i\) that is standardized with regard to the treatment effect standard error. Equation \eqref{borenstein} displays the general definition of \(Z^{(t)}\), the \(Z\)-score of the combined effect estimated by a common/fixed-effect meta-analysis on studies \(1\) up to and including \(t\) (adapted notation from \citet[p. 66]{borenstein2009introduction}):

\begin{equation} \label{borenstein}
\begin{split}
Z^{(t)} &= \frac{M^{(t)}}{SE_{M^{(t)}}} \\
M^{(t)} = \frac{\sum_{i = 1}^{t}W_{i}\overline{D}_{i}}{\sum_{i = 1}^{t}W_{i}} \quad W_{i} &= \frac{1}{SE^{2}_{\overline{D}_{i}}} \quad SE_{M^{(t)} }= \sqrt{\frac{1}{\sum_{i = 1}^{t}W_{i}}}
\end{split}
\end{equation}
Let \(d_{i} = \frac{\overline{D}_{i}}{\sigma_{D}} \) be the Cohen's \(d\) of the treatment score in study \(i\) \cite[p. 26]{borenstein2009introduction} --- so standardized with regard to the estimated population standard deviation --- and let \(n_{i}\) denote the sample size in the treatment and placebo arm of study \(i\) (under the assumption that all studies have equal size study arms). Since \(SE^{2}_{d_{i}} = \frac{1}{n_{i}}\), we let \(w_{i} = \frac{1}{SE^{2}_{d_{i}}} = \frac{1}{\frac{1}{n_{i}}} = n_{i}\) denote the weights for \(d_{i}\). Based on these weights, \(M^{(t)}\) and  \(SE_{M^{(t)}}\) can be expressed as follows, using the fact that \(\overline{D}_{i} = d_{i}\sigma_{D}\), \(SE^{2}_{\overline{D}_{i}} = \frac{\sigma^{2}_{D}}{n_{i}}\),  and thus \(W_{i} = w_{i}\frac{1}{\sigma^2_{D}}\) (see also \citet[p. 82]{borenstein2009introduction}):

\begin{equation}
\begin{split}
M^{(t)} &= \frac{\sum_{i = 1}^{t}w_{i}\frac{1}{\sigma^2_{D}}d_{i}\sigma_{D}}{\sum_{i = 1}^{t}w_{i}\frac{1}{\sigma^2_{D}}} = \frac{\sum_{i = 1}^{t}w_{i}d_{i}\sigma_{D}}{\sum_{i = 1}^{t}w_{i}}
= \frac{\sum_{i = 1}^{t}n_{i}d_{i}\sigma_{D}}{\sum_{i = 1}^{t}n_{i}}\\
SE_{M^{(t)}} &= \sqrt{\frac{1}{\sum_{i = 1}^{t}w_{i}\frac{1}{\sigma^2_{D}}}} = \sqrt{\frac{\sigma^{2}_{D}}{\sum_{i = 1}^{t}w_{i}}} = \sqrt{\frac{\sigma^{2}_{D}}{\sum_{i = 1}^{t}n_{i}}}
\end{split}
\end{equation}
With \(N^{(t)} = \sum_{i = 1}^{t} n_{i}\) and \(d_{i} = \frac{Z_{i}}{\sqrt{n_{i}}}\), the common/fixed-effect \(Z\)-score \(Z^{(t)}\) of studies \(i\) up to and including \(t\) can be derived as an average weighted by the square root of the individual study sample sizes:

\begin{equation} \label{Ztderiv}
\begin{split}
Z^{(t)} = \frac{\frac{\sum_{i = 1}^{t} n_{i}d_{i}\sigma_{D}}{N^{(t)}}}{\sqrt{\frac{\sigma^{2}_{D}}{N^{(t)}}}} 
= \frac{\sum_{i = 1}^{t} n_{i}d_{i}}{\sqrt{\sum_{i = 1}^{t} n_{i}}} 
= \frac{\sum_{i = 1}^{t} n_{i}\frac{Z_{i}}{\sqrt{n_{i}}}}{\sqrt{N^{(t)}}} 
&= \frac{\sum_{i = 1}^{t} \sqrt{n_{i}}Z_{i}}{\sqrt{N^{(t)}}} \\
&= \frac{\sum_{i = 1}^{t} \sqrt{n}Z_{i}}{\sqrt{t}\sqrt{n}} = \frac{1}{\sqrt{t}}\sum_{i = 1}^t Z_i \quad \text{for } n_1 = n_2 = \ldots = n_t = n
\end{split}
\end{equation}

\clearpage

\subsection{Expectation \lingform{Gold Rush} conditional pilot \(Z\)-score} \label{sec:AppExpPilot}
Here, and in the following, we assume that there is always a first study (\(\Prob\left[T\geq 1\right] = 1\)).

\begin{equation}
\begin{split}
\ExpNull\left[Z_1  \,\middle|\, T \geq 2\right] &= \frac{\ExpNull\left[Z_1  \,\middle|\, T \geq 2, Z_1 \geq z_{\frac{\alpha}{2}} \right] \cdot \ProbNull\left[T \geq 2 \,\middle|\, T \geq 1, Z_1 \geq z_{\frac{\alpha}{2}} \right] \cdot \ProbNull\left[Z_1 \geq z_{\frac{\alpha}{2}} \right]}
{\ProbNull\left[T \geq 2\right]}\\
& \quad +\frac{ \ExpNull\left[Z_1 \,\middle|\, T \geq 2,  \abs{Z_1} < z_{\frac{\alpha}{2}} \right] \cdot \ProbNull\left[T \geq 2 \,\middle|\, T \geq 1, \abs{Z_1} < z_{\frac{\alpha}{2}} \right] \cdot \ProbNull\left[\abs{Z_1} < z_{\frac{\alpha}{2}} \right]}
{\ProbNull\left[T \geq 2\right]} \\ \\
& = \frac{\ExpNull\left[Z_1  \,\middle|\, T \geq 2, Z_1 \geq z_{\frac{\alpha}{2}} \right] \cdot \omega_{\textsc{s}}^{(1)} \cdot \frac{\alpha}{2} + \ExpNull\left[Z_1  \,\middle|\, T \geq 2, \abs{Z_1} < z_{\frac{\alpha}{2}} \right] \cdot \omega_{\textsc{ns}}^{(1)} \cdot (1-\alpha)}{\omega_{\textsc{s}}^{(1)} \cdot \frac{\alpha}{2} + \omega_{\textsc{ns}}^{(1)} \cdot (1-\alpha)}
\end{split}
\end{equation}
\begin{equation}
\begin{split}
&\text{since} \\
\nonumber
\ProbNull\left[T \geq 2\right] &= \ProbNull\left[T \geq 2 \,\middle|\, T \geq 1, Z_1 \geq z_{\frac{\alpha}{2}}\right] \cdot \ProbNull\left[Z_1 \geq z_{\frac{\alpha}{2}}\right] 
+ \ProbNull\left[T \geq 2 \,\middle|\, T \geq 1, \abs{Z_1} < z_{\frac{\alpha}{2}}\right] \cdot \ProbNull\left[\abs{Z_1} < z_{\frac{\alpha}{2}}\right] \\
&= \omega_{\textsc{s}}^{(1)} \cdot \frac{\alpha}{2} + \omega_{\textsc{ns}}^{(1)} \cdot (1-\alpha)
\end{split}
\end{equation}
This expression only considers significant positive and nonsignificant results in the pilot study, since we defined in Eq. \eqref{newStudyProbs} that significant negative results have 0 probability to produce replication studies. We can replace \(\ProbNull\) by \(\Prob\) in the middle term of the fractions in the first two rows because \lingform{new study probabilities} are independent from the data generating distribution, as discussed in Section \ref{sec:GoldNewStudyProbsInd}.

\subsection{Expectation \lingform{Gold Rush} conditional meta-analysis \(Z\)-score} \label{sec:AppExpMeta}

\begin{equation}
\begin{split}
\text{For all } t \geq 2: \\
\ExpNull\left[Z^{(t)} \,\middle|\, T \geq t \right] 
 =& \frac{\sum_{i = 1}^{t} \sqrt{n_{i}} \ExpNull\left[Z_{i} \,\middle|\, T \geq t \right]}{\sqrt{N^{(t)}}} \\ \\
 =& \frac{\sqrt{n_{1}} \ExpNull\left[Z_1 \,\middle|\, T \geq t \right] + 
\sum_{i = 2}^{t-1} \sqrt{n_{i}} \ExpNull\left[Z_{i} \,\middle|\, T \geq t \right] + \sqrt{n_{t}} \ExpNull\left[Z_{t} \,\middle|\, T \geq t \right]}{\sqrt{N^{(t)}}} \\
 =& \frac{\sqrt{n_{1}} \ExpNull\left[Z_1 \,\middle|\, T \geq 2\right] + \sum_{i = 2}^{t-1} \sqrt{n_{i}} \ExpNull\left[Z_{i} \,\middle|\, T \geq i+1 \right]}{\sqrt{N^{(t)}}}
\end{split}
\end{equation}
Here we use that the last study in a series under the \lingform{Gold Rush} example is unbiased and has expectation 0 under the null hypothesis. We also use that the expansion of the series beyond the next study does not influence a study's expectation in our \lingform{Gold Rush} example: for \(t \geq 2\) \(\ExpNull\left[Z_1   \,\middle|\, T \geq t\right]\) is the same as \(\ExpNull\left[Z_1   \,\middle|\, T \geq 2\right]\), and for any \(i\) and \(t \geq i\), \(\ExpNull\left[Z_i   \,\middle|\, T \geq t\right]\) is the same as \(\ExpNull\left[Z_i   \,\middle|\, i + 1\right]\)).

\subsection{Mixture variance} \label{sec:AppMixVar}

\begin{subequations}
\begin{align}
\nonumber
&\Var\left\{  Z^{(2)}  \,\middle|\, T \geq 2  \right\} \\
\nonumber
& \qquad =  \frac{\alpha}{2} \cdot \omega_{\textsc{s}}^{(1)} \cdot \ExpNull\left[ \left( Z^{(2)} \right)^2  \,\middle|\, Z_1 \geq z_{\frac{\alpha}{2}} \right] 
+  (1 - \alpha) \cdot \omega_{\textsc{ns}}^{(1)} \cdot \ExpNull\left[ \left( Z^{(2)} \right)^2  \,\middle|\, \abs{Z_1} < z_{\frac{\alpha}{2}} \right]  \\
\nonumber
& \qquad \quad - \left( \frac{\alpha}{2} \cdot \omega_{\textsc{s}}^{(1)} \cdot \ExpNull\left[ Z^{(2)}  \,\middle|\, Z_1 \geq z_{\frac{\alpha}{2}} \right] 
+  (1 - \alpha) \cdot \omega_{\textsc{ns}}^{(1)} \cdot \ExpNull\left[ Z^{(2)}  \,\middle|\, \abs{Z_1} < z_{\frac{\alpha}{2}} \right]\right)^2 \\
\nonumber
& \qquad = \frac{\alpha}{2} \cdot \omega_{\textsc{s}}^{(1)} \cdot \left( \Var\left\{Z^{(2)}  \,\middle|\, Z_1 \geq z_{\frac{\alpha}{2}} \right\} + \ExpNull\left[  Z^{(2)}   \,\middle|\, Z_1 \geq z_{\frac{\alpha}{2}} \right]^2 \right) \\
\nonumber
& \qquad \quad +  (1 - \alpha) \cdot \omega_{\textsc{ns}}^{(1)} \cdot \left( \Var\left\{Z^{(2)}  \,\middle|\, \abs{Z_1} > z_{\frac{\alpha}{2}} \right\} + \ExpNull\left[  Z^{(2)}   \,\middle|\, Z_1 \geq z_{\frac{\alpha}{2}} \right]^2 \right)  \\
\nonumber
& \qquad \quad - \left( \frac{\alpha}{2} \cdot \omega_{\textsc{s}}^{(1)} \cdot \ExpNull\left[ Z^{(2)}  \,\middle|\, Z_1 \geq z_{\frac{\alpha}{2}} \right] 
+  (1 - \alpha) \cdot \omega_{\textsc{ns}}^{(1)} \cdot \ExpNull\left[ Z^{(2)}  \,\middle|\, \abs{Z_1} < z_{\frac{\alpha}{2}} \right]\right)^2 \\
\nonumber
& \qquad = \frac{\alpha}{2} \cdot \omega_{\textsc{s}}^{(1)} \cdot  \Var\left\{Z^{(2)}  \,\middle|\, Z_1 \geq z_{\frac{\alpha}{2}} \right\} 
 +  (1 - \alpha) \cdot \omega_{\textsc{ns}}^{(1)} \cdot \Var\left\{Z^{(2)}  \,\middle|\, \abs{Z_1} > z_{\frac{\alpha}{2}} \right\} \\
\label{averageSquaredMean}
& \qquad \quad + \frac{\alpha}{2} \cdot \omega_{\textsc{s}}^{(1)} \cdot \ExpNull\left[  Z^{(2)}   \,\middle|\, Z_1 \geq z_{\frac{\alpha}{2}} \right]^2  
+  (1 - \alpha) \cdot \omega_{\textsc{ns}}^{(1)} \cdot \ExpNull\left[  Z^{(2)}   \,\middle|\, Z_1 \geq z_{\frac{\alpha}{2}} \right]^2  \\
\label{squareAverageMean}
& \qquad \quad - \left( \frac{\alpha}{2} \cdot \omega_{\textsc{s}}^{(1)} \cdot \ExpNull\left[ Z^{(2)}  \,\middle|\, Z_1 \geq z_{\frac{\alpha}{2}} \right] 
+  (1 - \alpha) \cdot \omega_{\textsc{ns}}^{(1)} \cdot \ExpNull\left[ Z^{(2)}  \,\middle|\, \abs{Z_1} < z_{\frac{\alpha}{2}} \right]\right)^2
\end{align}
\end{subequations}
Because squaring is a convex function, we know from Jensen's Inequality that the average squared mean \eqref{averageSquaredMean} is larger than the square of the average mean \eqref{squareAverageMean}. So the variance of the mixture is larger than the mixture of the variances.

\subsection{Maximum time probability} \label{sec:AppMaxTime}

The survival function \(S(t - 1)\) represents the probability \(\Prob[T \geq t]\). The survival function is the complement of a cumulative distribution function on maximum time or stopping times T, known in survival analysis as the \lingform{lifetime distribution function} \(F(t - 1)\):
\begin{equation}
\begin{split}
S(t - 1) &= 1 - F(t - 1) \\
\text{with } \quad  F(t - 1) &= \sum_{i = 0}^{t-1} \Prob[T = i] 
\end{split}
\end{equation}

\begin{equation}
\begin{split}
S(t - 1) &= 1 - \sum_{i = 0}^{t  - 1} \Prob[T = i] \\
S(t) &= 1 -  \sum_{i = 0}^{t - 1} \Prob[T = i] - \Prob[T = t] \\
\text{therefore:} \quad \Prob[T = t] &= S(t - 1) - S(t)
\end{split}
\end{equation}

\subsection{Error control surviving over time in terms of a sum} \label{sec:AppFromSumToExists}
Let $\cF'^{(t)}_{\textsc{type-I}}$ be the even that both $\cF^{(t)}$ and $T \geq t$ holds. Using in the first equality below that the events $\cF'^{(1)}_{\textsc{type-I}}, \cF'^{(2)}_{\textsc{type-I}}, \ldots$ are all mutually exclusive (so that the union bound becomes an equality), we get: 
\begin{align*}
\sum_t \ProbNull\left[\cF^{(t)}_{\textsc{type-I}}, \cA^{(t)}, T \geq t  \right] &\leq
\sum_t \ProbNull\left[\cF^{(t)}_{\textsc{type-I}}, T \geq t \right] \\
& \quad = \ProbNull\left[\exists t > 0:\; \cF^{(t)}_{\textsc{type-I}}, T \geq t \right] \\
& \qquad \leq  \ProbNull\left[\exists t > 0:\; \cF^{(t)}_{\textsc{type-I}} \right] \\
& \qquad \quad = \ProbNull\left[\exists t > 0:\; \cE^{(t)}_{\textsc{type-I}} \right] \\
& \qquad \quad =  \ProbNull\left[\exists t > 0:\;  \LR^{(t)}\left(Z_1, \ldots, Z_t  \right) \geq \frac{1}{\alpha} 
\right] \leq \alpha
\end{align*}
where the final inequality is just the final inequality of (\ref{eq:lifeline}) again. 
(\ref{eq:deadline}) follows. 

\clearpage

\subsection{Code availability} \label{sec:AppCode}
Table \ref{tab:ExpZm}, Figure \ref{fig:distrZ} and Table \ref{tab:TypeIerrorTildeSim} were calculated, simulated and created by R code available in the EASY-DANS repository: \url{https://doi.org/10.17026/dans-x56-qfme} (see Extended data\cite{EASY-DANS}) 

Details on the OS and version at which it were run can be found below:

\begin{itemize}
\item Platform: x86 64-redhat-linux-gnu
\item Arch: x86 64
\item OS: linux-gnu
\item System: x86 64, linux-gnu
\item R version: 3.5.3 (2019-03-11) Great Truth
\item svn rev: 76217
\end{itemize}

The following packages were used:
\begin{itemize}
    \item ggplot2 version 3.0.0
    \item graphics version 3.5.3
    \item grDevices version 3.5.3
    \item methods version 3.5.3
    \item stats version 3.5.3
    \item utils version 3.5.3
\end{itemize}

%% file: main.bbl
\begin{thebibliography}{}

\bibitem[Armitage, 1984]{armitage1984controversies}
Armitage, P. (1984).
\newblock Controversies and achievements in clinical trials.
\newblock {\em Contemporary Clinical Trials}, 5(1):67--72.

\bibitem[Bayarri et~al., 2016]{bayarri2016rejection}
Bayarri, M., Benjamin, D.~J., Berger, J.~O., and Sellke, T.~M. (2016).
\newblock Rejection odds and rejection ratios: A proposal for statistical
  practice in testing hypotheses.
\newblock {\em Journal of Mathematical Psychology}, 72:90--103.

\bibitem[Berger and Berry, 1988]{berger1988relevance}
Berger, J.~O. and Berry, D.~A. (1988).
\newblock The relevance of stopping rules in statistical inference.
\newblock {\em Statistical decision theory and related topics IV}, 1:29--47.

\bibitem[Borenstein et~al., 2009]{borenstein2009introduction}
Borenstein, M., Hedges, L.~V., Higgins, J. P.~T., and Rothstein, H.~R. (2009).
\newblock {\em Introduction to Meta-Analysis}.
\newblock John Wiley \& Sons, Ltd.
\newblock {DOI}: 10.1002/9780470743386.refs.

\bibitem[Brok et~al., 2008]{brok2008apparently}
Brok, J., Thorlund, K., Wetterslev, J., and Gluud, C. (2008).
\newblock Apparently conclusive meta-analyses may be inconclusive—trial
  sequential analysis adjustment of random error risk due to repetitive testing
  of accumulating data in apparently conclusive neonatal meta-analyses.
\newblock {\em International journal of epidemiology}, 38(1):287--298.

\bibitem[Chalmers et~al., 2014]{chalmers2014increase}
Chalmers, I., Bracken, M.~B., Djulbegovic, B., Garattini, S., Grant, J.,
  G{\"u}lmezoglu, A.~M., Howells, D.~W., Ioannidis, J.~P., and Oliver, S.
  (2014).
\newblock How to increase value and reduce waste when research priorities are
  set.
\newblock {\em The Lancet}, 383(9912):156--165.

\bibitem[Chalmers and Glasziou, 2009]{chalmers2009avoidable}
Chalmers, I. and Glasziou, P. (2009).
\newblock Avoidable waste in the production and reporting of research evidence.
\newblock {\em The Lancet}, 114(6):1341--1345.

\bibitem[Chalmers and Glasziou, 2016]{chalmers2016systematic}
Chalmers, I. and Glasziou, P. (2016).
\newblock Systematic reviews and research waste.
\newblock {\em The Lancet}, 387(10014):122--123.

\bibitem[Chalmers and Lau, 1993]{chalmers1993meta}
Chalmers, T.~C. and Lau, J. (1993).
\newblock Meta-analytic stimulus for changes in clinical trials.
\newblock {\em Statistical Methods in Medical Research}, 2(2):161--172.

\bibitem[Claxton et~al., 2002]{claxton2002rational}
Claxton, K., Sculpher, M., and Drummond, M. (2002).
\newblock A rational framework for decision making by the national institute
  for clinical excellence ({N}{I}{C}{E}).
\newblock {\em The Lancet}, 360(9334):711--715.

\bibitem[Claxton and Sculpher, 2006]{claxton2006using}
Claxton, K.~P. and Sculpher, M.~J. (2006).
\newblock Using value of information analysis to prioritise health research.
\newblock {\em Pharmacoeconomics}, 24(11):1055--1068.

\bibitem[Davey et~al., 2011]{davey2011characteristics}
Davey, J., Turner, R.~M., Clarke, M.~J., and Higgins, J.~P. (2011).
\newblock Characteristics of meta-analyses and their component studies in the
  {C}ochrane database of systematic reviews: a cross-sectional, descriptive
  analysis.
\newblock {\em BMC medical research methodology}, 11(1):160.

\bibitem[Dawid, 1984]{dawid1984present}
Dawid, A.~P. (1984).
\newblock Present position and potential developments: Some personal views:
  statistical theory: the prequential approach.
\newblock {\em Journal of the Royal Statistical Society: Series A (General)},
  147(2):278--290.

\bibitem[Egger and Smith, 1998]{egger1998bias}
Egger, M. and Smith, G.~D. (1998).
\newblock Bias in location and selection of studies.
\newblock {\em BMJ: British Medical Journal}, 316(7124):61.

\bibitem[Ellis and Stewart, 2009]{ellis2009temporal}
Ellis, S.~P. and Stewart, J.~W. (2009).
\newblock Temporal dependence and bias in meta-analysis.
\newblock {\em Communications in Statistics—Theory and Methods},
  38(15):2453--2462.

\bibitem[Fergusson et~al., 2005]{fergusson2005randomized}
Fergusson, D., Glass, K.~C., Hutton, B., and Shapiro, S. (2005).
\newblock Randomized controlled trials of aprotinin in cardiac surgery: could
  clinical equipoise have stopped the bleeding?
\newblock {\em Clinical Trials}, 2(3):218--232.

\bibitem[Fisher, 1938]{fisher1938presidential}
Fisher, R.~A. (1938).
\newblock Presidential address.
\newblock {\em Sankhy{\=a}: The Indian Journal of Statistics}, pages 14--17.

\bibitem[Gehr et~al., 2006]{gehr2006fading}
Gehr, B.~T., Weiss, C., and Porzsolt, F. (2006).
\newblock The fading of reported effectiveness. a meta-analysis of randomised
  controlled trials.
\newblock {\em BMC medical research methodology}, 6(1):25.

\bibitem[G{\o}tzsche, 1987]{gotzsche1987reference}
G{\o}tzsche, P.~C. (1987).
\newblock Reference bias in reports of drug trials.
\newblock {\em Br Med J (Clin Res Ed)}, 295(6599):654--656.

\bibitem[Gr\"unwald et~al., 2019]{GrunwaldHK19}
Gr\"unwald, P.~D., De~Heide, R., and Koolen, W. (2019).
\newblock Safe testing.
\newblock {\em arXiv preprint}.

\bibitem[Hendriksen et~al., 2018]{hendriksen2018optional}
Hendriksen, A., de~Heide, R., and Gr{\"u}nwald, P. (2018).
\newblock Optional stopping with {B}ayes factors: a categorization and
  extension of folklore results, with an application to invariant situations.
\newblock {\em arXiv preprint arXiv:1807.09077}.

\bibitem[Higgins et~al., 2011]{higgins2011sequential}
Higgins, J., Whitehead, A., and Simmonds, M. (2011).
\newblock Sequential methods for random-effects meta-analysis.
\newblock {\em Statistics in medicine}, 30(9):903--921.

\bibitem[Ioannidis, 2010]{ioannidis2010meta}
Ioannidis, J. (2010).
\newblock Meta-research: The art of getting it wrong.
\newblock {\em Research Synthesis Methods}, 1(3-4):169--184.

\bibitem[Ioannidis, 2005a]{ioannidis2005contradicted}
Ioannidis, J.~P. (2005a).
\newblock Contradicted and initially stronger effects in highly cited clinical
  research.
\newblock {\em Jama}, 294(2):218--228.

\bibitem[Ioannidis, 2005b]{ioannidis2005most}
Ioannidis, J.~P. (2005b).
\newblock Why most published research findings are false.
\newblock {\em PLoS medicine}, 2(8):e124.

\bibitem[Ioannidis, 2008]{ioannidis2008most}
Ioannidis, J.~P. (2008).
\newblock Why most discovered true associations are inflated.
\newblock {\em Epidemiology}, pages 640--648.

\bibitem[Ioannidis and Trikalinos, 2005]{ioannidis2005early}
Ioannidis, J.~P. and Trikalinos, T.~A. (2005).
\newblock Early extreme contradictory estimates may appear in published
  research: the {P}roteus phenomenon in molecular genetics research and
  randomized trials.
\newblock {\em Journal of clinical epidemiology}, 58(6):543--549.

\bibitem[Krum and Tonkin, 2003]{krum2003phase}
Krum, H. and Tonkin, A. (2003).
\newblock Why do phase {III} trials of promising heart failure drugs often
  fail? the contribution of “regression to the truth”.
\newblock {\em Journal of cardiac failure}, 9(5):364--367.

\bibitem[Kulinskaya et~al., 2016]{kulinskaya2016sequential}
Kulinskaya, E., Huggins, R., and Dogo, S.~H. (2016).
\newblock Sequential biases in accumulating evidence.
\newblock {\em Research synthesis methods}, 7(3):294--305.

\bibitem[Lund et~al., 2016]{lund2016towards}
Lund, H., Brunnhuber, K., Juhl, C., Robinson, K., Leenaars, M., Dorch, B.~F.,
  Jamtvedt, G., Nortvedt, M.~W., Christensen, R., and Chalmers, I. (2016).
\newblock Towards evidence based research.
\newblock {\em Bmj}, 355:i5440.

\bibitem[Mallett and Clarke, 2003]{mallett2003many}
Mallett, S. and Clarke, M. (2003).
\newblock How many {C}ochrane reviews are needed to cover existing evidence on
  the effects of health care interventions?
\newblock {\em ACP journal club}, 139(1):A11--A11.

\bibitem[Moher et~al., 2007a]{moher2007epidemiology}
Moher, D., Tetzlaff, J., Tricco, A.~C., Sampson, M., and Altman, D.~G. (2007a).
\newblock Epidemiology and reporting characteristics of systematic reviews.
\newblock {\em PLoS medicine}, 4(3):e78.

\bibitem[Moher and Tsertsvadze, 2006]{moher2006systematic}
Moher, D. and Tsertsvadze, A. (2006).
\newblock Systematic reviews: when is an update an update?
\newblock {\em The Lancet}, 367(9514):881--883.

\bibitem[Moher et~al., 2008]{moher2008and}
Moher, D., Tsertsvadze, A., Tricco, A., Eccles, M., Grimshaw, J., Sampson, M.,
  and Barrowman, N. (2008).
\newblock When and how to update systematic reviews.
\newblock {\em Cochrane database of systematic reviews}, (1).

\bibitem[Moher et~al., 2007b]{moher2007systematic}
Moher, D., Tsertsvadze, A., Tricco, A.~C., Eccles, M., Grimshaw, J., Sampson,
  M., and Barrowman, N. (2007b).
\newblock A systematic review identified few methods and strategies describing
  when and how to update systematic reviews.
\newblock {\em Journal of clinical epidemiology}, 60(11):1095--e1.

\bibitem[Page et~al., 2016]{page2016epidemiology}
Page, M.~J., Shamseer, L., Altman, D.~G., Tetzlaff, J., Sampson, M., Tricco,
  A.~C., Catal{\'a}-L{\'o}pez, F., Li, L., Reid, E.~K., Sarkis-Onofre, R.,
  et~al. (2016).
\newblock Epidemiology and reporting characteristics of systematic reviews of
  biomedical research: a cross-sectional study.
\newblock {\em PLoS medicine}, 13(5):e1002028.

\bibitem[Pereira and Ioannidis, 2011]{pereira2011statistically}
Pereira, T.~V. and Ioannidis, J.~P. (2011).
\newblock Statistically significant meta-analyses of clinical trials have
  modest credibility and inflated effects.
\newblock {\em Journal of clinical epidemiology}, 64(10):1060--1069.

\bibitem[Pfeiffer et~al., 2011]{pfeiffer2011quantifying}
Pfeiffer, T., Bertram, L., and Ioannidis, J.~P. (2011).
\newblock Quantifying selective reporting and the {P}roteus phenomenon for
  multiple datasets with similar bias.
\newblock {\em PLoS One}, 6(3):e18362.

\bibitem[Proschan et~al., 2006]{proschan2006statistical}
Proschan, M.~A., Lan, K.~G., and Wittes, J.~T. (2006).
\newblock {\em Statistical monitoring of clinical trials: a unified approach}.
\newblock Springer Science \& Business Media.

\bibitem[Robbins, 1970]{robbins1970statistical}
Robbins, H. (1970).
\newblock Statistical methods related to the law of the iterated logarithm.
\newblock {\em Annals of Mathematical Statistics}, 41:1397--1409.

\bibitem[Roberts and Ker, 2015]{roberts2015systematic}
Roberts, I. and Ker, K. (2015).
\newblock How systematic reviews cause research waste.
\newblock {\em The Lancet}, 386(10003):1536.

\bibitem[Robinson and Goodman, 2011]{robinson2011systematic}
Robinson, K.~A. and Goodman, S.~N. (2011).
\newblock A systematic examination of the citation of prior research in reports
  of randomized, controlled trials.
\newblock {\em Annals of internal medicine}, 154(1):50--55.

\bibitem[Rosenthal, 1979]{rosenthal1979file}
Rosenthal, R. (1979).
\newblock The file drawer problem and tolerance for null results.
\newblock {\em Psychological bulletin}, 86(3):638.

\bibitem[Royall, 2000]{royall2000probability}
Royall, R. (2000).
\newblock On the probability of observing misleading statistical evidence.
\newblock {\em Journal of the American Statistical Association},
  95(451):760--768.

\bibitem[Schure, 2019]{EASY-DANS}
Schure, J.~T. (2019).
\newblock Accumulation bias in meta-analysis: The need to consider time in
  error control.

\bibitem[Shafer, 2019]{shafer2019language}
Shafer, G. (2019).
\newblock The language of betting as a strategy for statistical and scientific
  communication.
\newblock \url{http://probabilityandfinance.com/articles/54.pdf}.
\newblock Online; accessed 16 May 2019.

\bibitem[Shafer et~al., 2011]{shafer2011test}
Shafer, G., Shen, A., Vereshchagin, N., Vovk, V., et~al. (2011).
\newblock Test martingales, {B}ayes factors and p-values.
\newblock {\em Statistical Science}, 26(1):84--101.

\bibitem[Shafer and Vovk, 2019]{shafer2019game}
Shafer, G. and Vovk, V. (2019).
\newblock {\em Game-Theoretic Foundations for Probability and Finance}.
\newblock Wiley.

\bibitem[Simmonds et~al., 2017]{simmonds2017living}
Simmonds, M., Salanti, G., McKenzie, J., and Elliott, J. (2017).
\newblock Living systematic reviews: 3. statistical methods for updating
  meta-analyses.
\newblock {\em Journal of clinical epidemiology}, 91:38--46.

\bibitem[Thorlund et~al., 2008]{thorlund2008can}
Thorlund, K., Devereaux, P., Wetterslev, J., Guyatt, G., Ioannidis, J.~P.,
  Thabane, L., Gluud, L.-L., Als-Nielsen, B., and Gluud, C. (2008).
\newblock Can trial sequential monitoring boundaries reduce spurious inferences
  from meta-analyses?
\newblock {\em International journal of epidemiology}, 38(1):276--286.

\bibitem[Turner et~al., 2013]{turner2013impact}
Turner, R.~M., Bird, S.~M., and Higgins, J.~P. (2013).
\newblock The impact of study size on meta-analyses: examination of
  underpowered studies in {C}ochrane reviews.
\newblock {\em PloS one}, 8(3):e59202.

\bibitem[Wetterslev et~al., 2017]{wetterslev2017trial}
Wetterslev, J., Jakobsen, J.~C., and Gluud, C. (2017).
\newblock Trial sequential analysis in systematic reviews with meta-analysis.
\newblock {\em BMC medical research methodology}, 17(1):39.

\bibitem[Wetterslev et~al., 2008]{wetterslev2008trial}
Wetterslev, J., Thorlund, K., Brok, J., and Gluud, C. (2008).
\newblock Trial sequential analysis may establish when firm evidence is reached
  in cumulative meta-analysis.
\newblock {\em Journal of clinical epidemiology}, 61(1):64--75.

\bibitem[Whitehead, 1997]{whitehead1997prospectively}
Whitehead, A. (1997).
\newblock A prospectively planned cumulative meta-analysis applied to a series
  of concurrent clinical trials.
\newblock {\em Statistics in medicine}, 16(24):2901--2913.

\bibitem[Whitehead, 2002]{whitehead2002meta}
Whitehead, A. (2002).
\newblock {\em Meta-analysis of controlled clinical trials}, volume~7.
\newblock John Wiley \& Sons.

\end{thebibliography}
